\documentclass[fleqn,usenatbib]{mnras}
\usepackage{newtxtext,newtxmath}
\usepackage[table,xcdraw]{xcolor}
\usepackage[T1]{fontenc}
\usepackage{tabularx}
\usepackage[table,xcdraw]{xcolor}
\usepackage{gensymb}
\usepackage{hyperref}
\hypersetup{
    colorlinks=true,
    linkcolor=blue,
    filecolor=magenta,      
    urlcolor=red,
    pdfpagemode=FullScreen,
    }
\usepackage{multirow}

\DeclareRobustCommand{\VAN}[3]{#2}
\let\VANthebibliography\thebibliography
\def\thebibliography{\DeclareRobustCommand{\VAN}[3]{##3}\VANthebibliography}

\usepackage{graphicx}	
\usepackage{amsmath}	
\usepackage{amsfonts}

\usepackage{float}
\usepackage[caption = false]{subfig}
\usepackage{comment}

\usepackage{siunitx}

\usepackage{tikz,xcolor,hyperref}
\definecolor{lime}{HTML}{A6CE39}
\DeclareRobustCommand{\orcidicon}{%
	\begin{tikzpicture}
	\draw[lime, fill=lime] (0,0) 
	circle [radius=0.16] 
	node[white] {{\fontfamily{qag}\selectfont \tiny ID}};
	\draw[white, fill=white] (-0.0625,0.095) 
	circle [radius=0.007];
	\end{tikzpicture}
	\hspace{-2mm}
}
\foreach \x in {A, ..., Z}{%
	\expandafter\xdef\csname orcid\x\endcsname{\noexpand\href{https://orcid.org/\csname orcidauthor\x\endcsname}{\noexpand\orcidicon}}
}

\title[PKS 0346-27 ]{Multi-frequency Variability Study of Flat-Spectrum Radio Quasar PKS 0346-27}

\author[
Kamaram, Prince, Pramanick \& Bose]{
Sushanth Reddy Kamaram\orcidB $^{1}$,
Raj Prince\orcidC$^{2}$,
Suman Pramanick\orcidA$^{1}$ 
and Debanjan Bose\orcidD $^{3}$ \thanks{E-mail: debanjan.bose@bose.res.in}
\\
$^{1}$Department of Physics, Indian Institute of Technology Kharagpur, 721302, India\\
$^{2}$Center for Theoretical Physics, Polish Academy of Sciences, Lotnik\'{o}w 32/46, Warsaw, Poland\\
$^{3}$Department of Astrophysics and High Energy Physics, S N Bose National Centre for Basic Sciences, Kolkata, 700106, India
}

\date{Accepted XXX. Received YYY; in original form ZZZ}

\pubyear{2022}

\begin{document}
\label{firstpage}
\pagerange{\pageref{firstpage}--\pageref{lastpage}}

\maketitle

\begin{abstract}
We have presented a multiwavelength temporal and spectral study of the Blazar PKS 0346-27 for the period 2019 January–2021 December (MJD 58484–59575)  using data from Fermi-LAT ($\gamma$-rays), Swift-XRT (X-rays) and Swift-UVOT (ultra-violet and optical).  We identified multiple flaring episodes by analyzing the gamma-ray light curve generated from the Fermi-LAT data over a two-year period. The light curves of these individual gamma-ray flares with one-day binning were then modeled using a sum-of-exponentials fit. We found the minimum variability times for the gamma-ray light curve to be $1.34 \pm 0.3$ days and a range of 0.1–3.2 days for the Swift wavelengths suggesting the compactness of the source. The broadband emission mechanism was studied by modeling the simultaneous multi-waveband Spectral Energy Distributions (SED) using  the one-zone leptonic emission mechanism. We found that the optical-UV and X-ray data can be explained by the synchrotron and Synchrotron Self-Compton (SSC) emissions. However, the disk component of the External Compton radiation is dominant at higher energies with contributions from the EC broad line region component and SSC. Further, we performed a power spectral density (PSD) analysis with data from the gamma-ray light curve using the power spectrum response (PSRESP) method.  With the power law model, a best-fit slope of $2.15 \pm 0.87$ was found. This source could be a promising target for upcoming CTA for its harder spectrum at lower energies (tens of GeV).

\end{abstract}



\section{Introduction}
Radio-loud Active Galactic Nuclei (AGN) are the most commonly observed astrophysical sources of high energy. They emit bright non-thermal radiation across a wide range of wavelengths (from radio to very high energy $\gamma$-rays) through relativistic jets assumed to be produced by the  interaction of strong magnetic field lines and ionized particles accelerated close to the speed of light with a supermassive black hole (SMBH) at the center. If the relativistic jet of the AGN is aligned at a small angle with the observer's line of sight (LOS) then it is called a blazar (\cite{urry1995unified}). In most of the blazars,  we observe a wide range of variability times across the electromagnetic spectrum (\cite{aharonian2007exceptional}; \cite{raiteri2013awakening}). Historically, blazars are classified into two types depending on their optical spectra: the BL Lacertae objects (BL Lacs) typically have very weak or absent features in their optical spectra, and the Flat-Spectrum Radio Quasars (FSRQs)  show strong broad optical emission lines. \cite{berton2018radio} show that the flat-spectrum radio-loud Narrow-Line Seyfert 1 (NLS1) galaxies can host low-power blazar-type jets aligned with our Los. The observed broadband spectral energy distributions (SEDs) of blazars show a typical double-peaked structure. The low energy peak, extending from radio to optical wavelengths, is produced by the synchrotron emissions of relativistic electrons accelerated by the jet magnetic field.  Blazars are also divided into three main classes depending on the position of their low energy or synchrotron peak. If the synchrotron peak is observed at $<10^{14}Hz$, then it is called a low synchrotron peak (LSP), if the synchrotron peak is observed between $10^{14}Hz$ and $10^{15}Hz$, then they are called intermediate synchrotron peaked (ISP) and finally, synchrotron peak $\geq 10^{15}Hz$ are called high synchrotron peaked (HSP) blazar \citep{abdo2010spectral}. The emission processes that give rise to the high-energy peak are not clear. In the simplest leptonic emission model the high energy $\gamma$-ray radiation results from the Inverse-Compton (IC) scattering of soft target photons originating in the synchrotron radiation process (SSC; \cite{sikora2009constraining}) or external photon fields (EC; \cite{dermer1992high}; \cite{sikora1994comptonization}). More advanced models incorporate relativistic protons as well and are called lepto-hadronic emission models (e.g. \cite{bottcher2013leptonic}).

PKS 0346-27 is an FSRQ class blazar with coordinates R.A. = $57.1589354^{\circ}$, Decl. = $-27.8204344^{\circ}$ (J2000, \cite{beasley2002vlba}) and a redshift, z = 0.991 (\cite{white1988redshifts}). It is also known as BZQ J0348-2749 ( \cite{massaro2009roma}). In the Parkes catalog (\cite{bolton1964parkes}), it was first identified as a radio source. Based on its optical spectrum, it was initially classified as a quasar (\cite{white1988redshifts}). Later it was revealed as an X-ray source by ROSAT (\cite{voges1999rosat} and references therein). Energetic Gamma Ray Experiment Telescope (EGRET, \cite{thompson1993calibration})  has not detected any $\gamma$-ray emissions from this source. However,  this changed when $\gamma$-rays were first detected from PKS 0346-27 region by Fermi-LAT and  the source was included in the Fermi-LAT First Source Catalog (1FGL, \cite{abdo2010fermi}). In the latest Fermi-LAT Fourth Source Catalog (4FGL, \cite{lat2019fermi}), it is associated with the $\gamma$-ray source 4FGL J0348.5-2749.

Based on the data taken on 2017 Nov 14 (MJD 58071), a near-infrared (NIR) flare was first reported from PKS 0346-27 (\cite{2017ATel10999....1C}). A few months later, on 2018 Feb 02 (MJD 58151), strong $\gamma$-ray flaring activity was reported from the source based on the Fermi-LAT data (\cite{angioni2018fermi}). PKS 0346-27 was found in an elevated state with a significantly harder spectrum with respect to the one reported in the 3FGL and reaching a daily $\gamma$-ray flux (energy > 100 MeV) more than 100 times larger than the average flux reported in the 3FGL. Multiwavelength follow-up observations revealed enhanced activity in the optical-NIR (\cite{nesci2018high}; \cite{vallely2018asas}), ultra-violet (UV) and X-ray (\cite{nesci2018x}). A multiwavelength broad-band spectral study has been reported by \cite{angioni2019large}.  They did a comparative study between  the quiescent and the flaring states in 2018. They fitted the SEDs with a one-zone leptonic model and found that the gamma-ray emission region has a lower magnetic field and a higher Lorentz factor for the flaring state in comparison to a quiescent state. They also found that the peak position of the low-energy hump shifts as the source goes into a flaring state towards higher energies, meaning the blazar classification mentioned earlier is time-dependent. During the quiescent state, the source would be classified as an LSP, and during the flaring state as an ISP. In our analysis, since we have analyzed the source only when it is in a flaring state, we do not see a change in the position of the low-energy peak. Peak positions for all flares above $10^{14}Hz$ meaning they would be classified as ISP.

In this work, we have studied the flaring states between 2019 January and 2021 December (MJD 58484-59575) using $\gamma$-ray and X-ray/UVOT data. We characterize the flaring activity of PKS 0346-27 for this period, probing the temporal behavior in $\gamma$-rays. Multiwavelength light curves are generated to identify flaring episodes. In section \ref{sec2}, we discuss multiwavelength observations and data analysis techniques for different archival data. The sum-of-exponentials fitting of $\gamma$-ray flares is discussed in section \ref{sec3}. In section \ref{sec4}, the broadband spectral energy distributions (SEDs) and their modeling are discussed. The power density spectrum of $\gamma$-ray light curve is discussed in section \ref{sec5} and our results are discussed in section \ref{res_dis}. 

\section{Multiwavelength Observations and Data Analysis}\label{sec2}
\subsection{Fermi-LAT}
\label{subsec:fermi_lat}
Fermi is a space-based gamma-ray telescope launched in June 2008. The Large Area Telescope (LAT) and the Gamma-ray Burst Monitor (GBM) are the two instruments onboard the telescope. With a maximum effective area (1-10 Gev energy range) of $9500 \; cm^2$ at normal incidence, LAT mainly operates in the 20 Mev-300 GeV energy range, although it is sensitive to energies outside this interval. The on-axis energy resolution is 9\%-15\% for 100 MeV-1 GeV gamma-rays. The multiple interleaved tungsten and silicon layers act as a converter-tracker system. The tungsten layers convert the gamma-rays into electron-positron pairs and the silicon-strip layers record the tracks of these charged particles to estimate the direction of the incident radiation. The energy of an incident gamma-ray is estimated from the energy deposited by the corresponding electron-positron pair on the calorimeter made of Cs(Tl) crystals at the base of LAT. The instrument has a Field of View (FoV) of 2.4 sr along with an angular resolution of < 3.5\degree at 100 MeV and < 0.15\degree at 10 GeV (\cite{atwood2009large}). The cosmic rays are a major contributor to the background noise in the LAT operating energy range. The ratio of charged cosmic rays to gamma-ray detections for the LAT is in the range of $10^3 - 10^5$ (\cite{ajello2021fermi}). The Anti-Coincidence Detector (ACD) which surrounds the converter-tracker layers is responsible for differentiating between the two radiations.

Observations of the object PKS 0346-27 from MJD 58484 to MJD 59575 were selected for analysis with a $10\degree$ circular Region of Interest (ROI) centered at the source. (RA=57.1589, Dec=-27.8204).

The Fermi-LAT data were analyzed using the recommended fermitools\footnote{\url{https://github.com/fermi-lat/Fermitools-conda/}} package. The data in the energy range 100 MeV to 300 GeV was obtained for the period MJD 58484-59575. The photon data was filtered using \textit{gtselect} tool with the constraints \textit{evclass=128} and \textit{evtype=3}. The zenith angle for the observations was restricted to less than 90 degrees to avoid contamination from Earth's limb. The time intervals were filtered using the constraint ‘$(DATA\_QUAL>0)\&\&(LAT\_CONFIG==1)$’ in the \textit{gtmktime} tool. The light curve was binned in 1-day intervals using the \textit{gtbin} tool and the exposure in $cm^2 s$ was computed using \textit{gtexposure}. The flux can be obtained from the counts and exposure terms.

\begin{figure*}
    \centering
    \includegraphics[width=\textwidth]{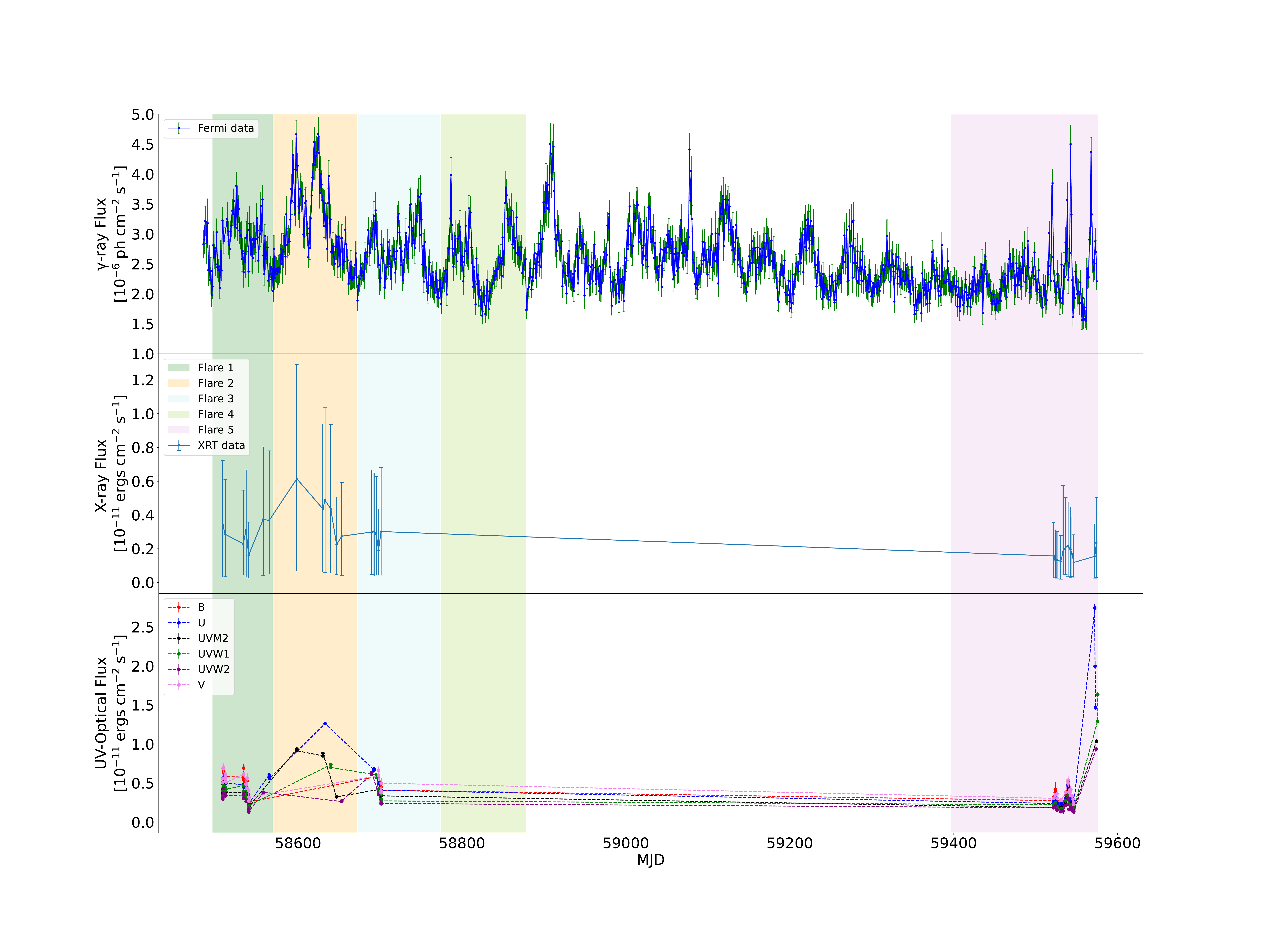}
    \caption{Multi-wavelength light curve of PKS 0346-27 in the period MJD 58484-59575. The time period is further divided into Flares 1-5 for observing changes in spectral parameters over time. (Top panel) One-day binned $\mathrm{\gamma-ray}$ light curve obtained from Fermi-LAT data using aperture photometry method in the 0.1-300 GeV energy range. (Middle panel) X-ray light curve generated from 30 Swift-XRT observations in the energy range 0.3-8 keV. (Bottom panel) Light curves for six Swift-UVOT bands (V, B, U, UVW1, UVM2, and UVW2) were obtained from the same observations as Swift-XRT. Swift data for the period between MJD 58702 and MJD 59520 is not available.} 
    \label{fig_broadband_lc}
\end{figure*}

Using the event file obtained in the gamma-ray light curve analysis, a live time HealPix table was computed for the region of interest from the \textit{expCube} tool. An exposure map for a radius of 10 degrees more than the ROI was computed using the live time HealPix table and the \textit{expMap} function with the instrument response file \textit{P8R3\_SOURCE\_V3} as input. A total of 131 point sources and 1 extended source were identified in the region of interest using the user-contributed tool \textit{make4FGLxml.py}\footnote{\url{https://fermi.gsfc.nasa.gov/ssc/data/analysis/user/make4FGLxml.py}}. The background files \textit{gll\_iem\_v07.fits}\footnote{\url{https://fermi.gsfc.nasa.gov/ssc/data/access/lat/BackgroundModels.html}} and \textit{iso\_P8R3\_SOURCE\_V3\_v1.txt}\footnote{\url{https://fermi.gsfc.nasa.gov/ssc/data/access/lat/BackgroundModels.html}} were used for the response model. The models and their respective parameters of the nearby sources were stored in an \textit{XML} file and the diffusion response was computed. The $\gamma$-ray light curve shown in the top panel of Figure \ref{fig_broadband_lc} is obtained with a 24-hour bin size. The gamma-ray spectral analysis is explained in section \ref{sec:gamma_ray_SED}.

\begin{figure}
    \centering
    \includegraphics[width=0.5\textwidth]{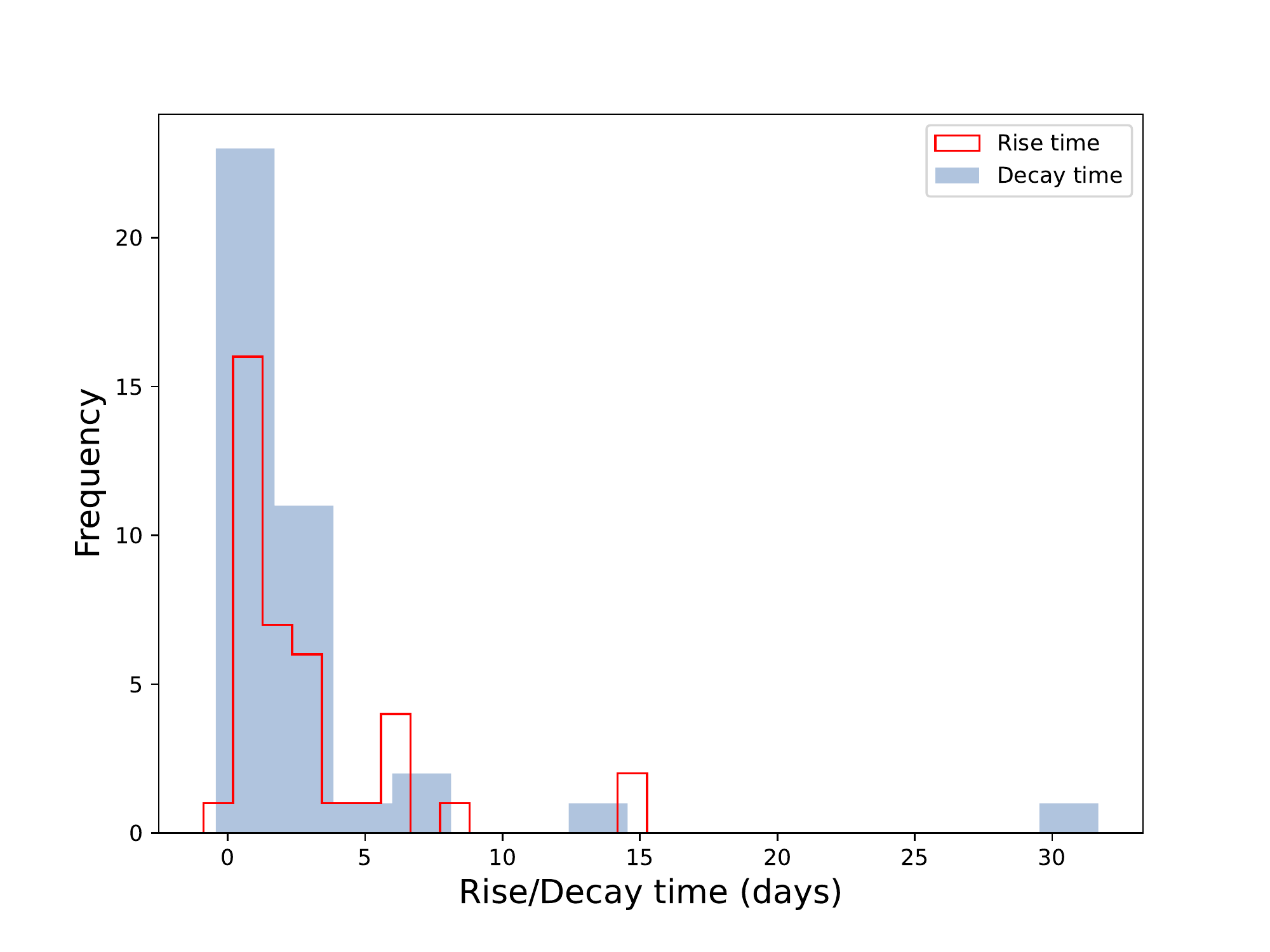}
    \caption{Statistical distribution of Rise ($\mathrm{R_t}$) and Decay ($\mathrm{D_t}$) times of the local peaks identified in the $\mathrm{\gamma-ray}$ light curve show in Figure \ref{fig_gamma_lc_fit}. }
    \label{fig:rise_and_decay_time_distribution}
\end{figure}

\subsection{Swift-XRT}
The Neil Gehrels Swift observatory was launched in November 2004 aboard the Delta II rocket with three instruments onboard i.e., the Burst Alert Telescope (BAT), the X-Ray Telescope (XRT), and the Ultra-violet Optical Telescope (UVOT). The Swift-XRT is a grazing incidence Wolter-I telescope with an effective area of 110 $cm^2$ and a 0.2-10 keV energy range.

A total of 30 observations from Swift-XRT (see Table \ref{tab_observation_ids}) were available in the same time period as the Fermi observations. The exposure times for these observations span from 1100 to 3800 seconds in the period MJD 58508 to MJD 58701. The lack of observations from MJD 58701 for X-ray analysis is a drawback. Software tools from the HEASoft package were used for analysis.

The XRT data was analyzed using Xselect and Xspec tools available in HEASoft Software.  The Level 1 data is passed to xrtpipeline resulting in the generation of Level 2 files used for further analysis. Using the event file and Xselect tools, the image file is extracted which is further used to create source and background regions in the DS9 viewer. The source region file contains a circular 60-arcsec region and the background region file is a circular 120-arcsec region adjacent to the source. Again using Xselect, both the light curves and the spectra were extracted for source and background regions.

The auxiliary response file was generated using the "xrtmkarf" tool, with the image and the source spectrum files generated using "Xselect" as inputs. The "quzcif" tool was used to find the corresponding redistribution matrix file. Using "grppha", the source and background files, the redistribution matrix file, and the auxiliary response file are combined into one spectrum file for Xspec analysis. Xspec was used to model the X-ray spectra with an energy range of 0.3-0.8 KeV. The power law model with $N_H (= 8.16 \times 10^{19} cm^{-2})$ was used to model the spectra. 

\subsection{Swift-UVOT}
In the same period as Swift-XRT observations, a total of 166 observations were available for Swift-UVOT analysis with exposure times ranging from 31 to 1626 seconds. The image data files for all observations were combined using the \textit{uvotimsum} tool. Then, the source and background files from the X-ray analysis were used as inputs to the \textit{uvotsource} tool along with the combined image file. The UVOT data is recorded at six wavelengths namely UVW2 (\SI{1928}{\angstrom}), UVM2 (\SI{2246}{\angstrom}), UVW1 (\SI{2600}{\angstrom}), V (\SI{3464}{\angstrom}), U (\SI{4392}{\angstrom}), B (\SI{5468}{\angstrom}). The flux obtained from "uvotsource" is multiplied with the corresponding wavelengths to obtain the energy flux in $\mathrm{erg \; cm^{-2} \; s^{-1}}$.

\begin{table}
\centering
\begin{tabular}{c c c}
\hline
\textbf{Observation ID} & \textbf{MJD} & \textbf{Exposure time (s)} \\ \hline
38373017                & 58508        & 3502.7                     \\
38373018                & 58511        & 3768.5                     \\
38373019                & 58533        & 1526.9                     \\
38373020                & 58537        & 2630.3                     \\
38373021                & 58540        & 2576.6                     \\
38373022                & 58557        & 2998.4                     \\
38373023                & 58565        & 2601.0                     \\
38373024                & 58598        & 2877.3                     \\
38373026                & 58630        & 1969.7                     \\
38373027                & 58633        & 2009.7                     \\
38373028                & 58640        & 1983.3                     \\
38373029                & 58647        & 1108.4                     \\
38373030                & 58653        & 1947.8                     \\
38373031                & 58690        & 1988.3                     \\
38373032                & 58693        & 1835.3                     \\
38373033                & 58695        & 1785.2                     \\
38373034                & 58698        & 2020.9                     \\
38373035                & 58701        & 1898.0                     \\
38373036                & 59522        & 1978.3                     \\
38373037                & 59524        & 1842.9                     \\
38373040                & 59526        & 2203.9                     \\
38373043                & 59530        & 2146.2                     \\
38373044                & 59533        & 569.2                      \\
38373045                & 59537        & 1609.7                     \\
38373046                & 59540        & 2101.1                     \\
38373047                & 59543        & 2492.3                     \\
38373048                & 59544        & 2173.8                     \\
38373050                & 59546        & 1860.4                     \\
38373051                & 59572        & 2028.4                     \\
38373052                & 59574        & 2733.0                     \\ \hline
\end{tabular}

\caption{Swift observations used for this analysis}
\label{tab_observation_ids}
\end{table}

\section{Multi-waveband light curve}\label{sec3}
\subsection{Fitting the light curve}
The gamma-ray light curve with 24 hr bins was divided into five flares corresponding to the largest peaks and troughs. The segments flare 1, flare 2, flare 3, flare 4, and flare 5 have the time periods MJD 58494-58569, MJD 58570-58672, MJD 58673-58774, MJD 58775-58879, and MJD 59396-59575 respectively. Further, it is observed that individual flares have their own substructure with multiple peaks (see Figure \ref{fig_gamma_lc_fit}). These are modeled using the sum-of-exponentials function given in equation \ref{eqn:sum_of_exponetials}. 

\begin{figure*}
    \centering
    \subfloat[]{\includegraphics[width = 0.5 \textwidth]{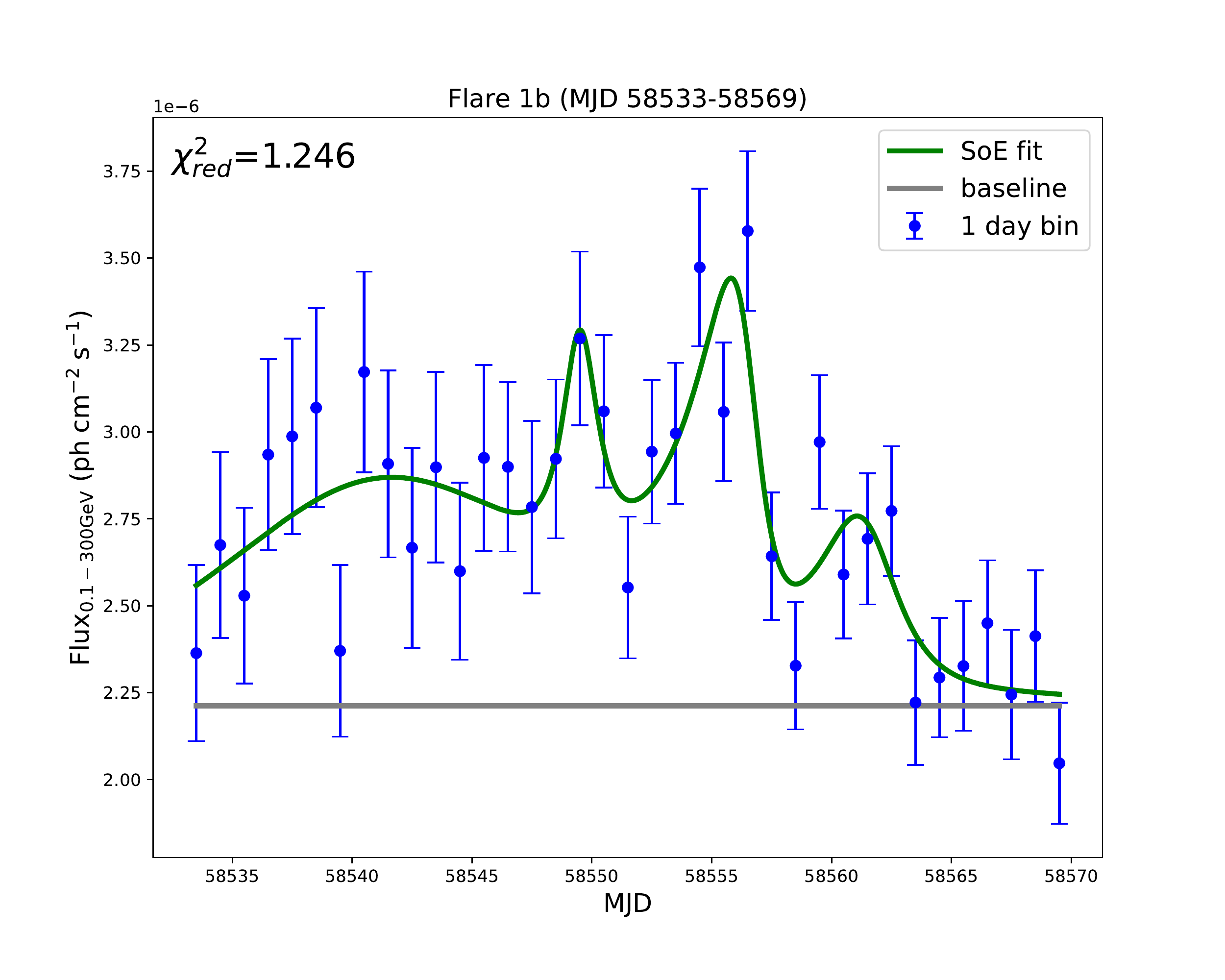}}\\[-5ex]
    \subfloat[]{\includegraphics[width = 0.5 \textwidth]{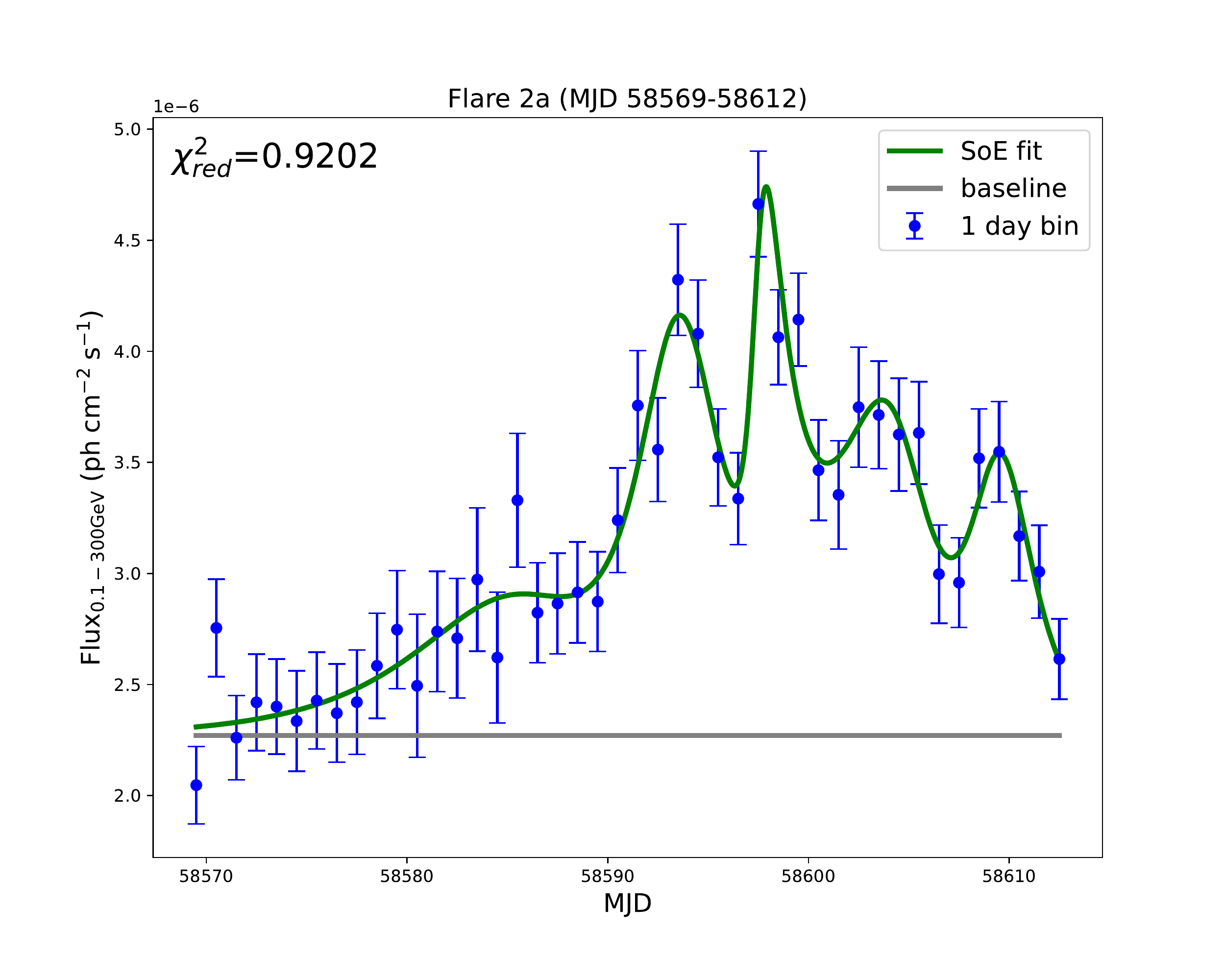}} 
    \subfloat[]{\includegraphics[width = 0.5 \textwidth]{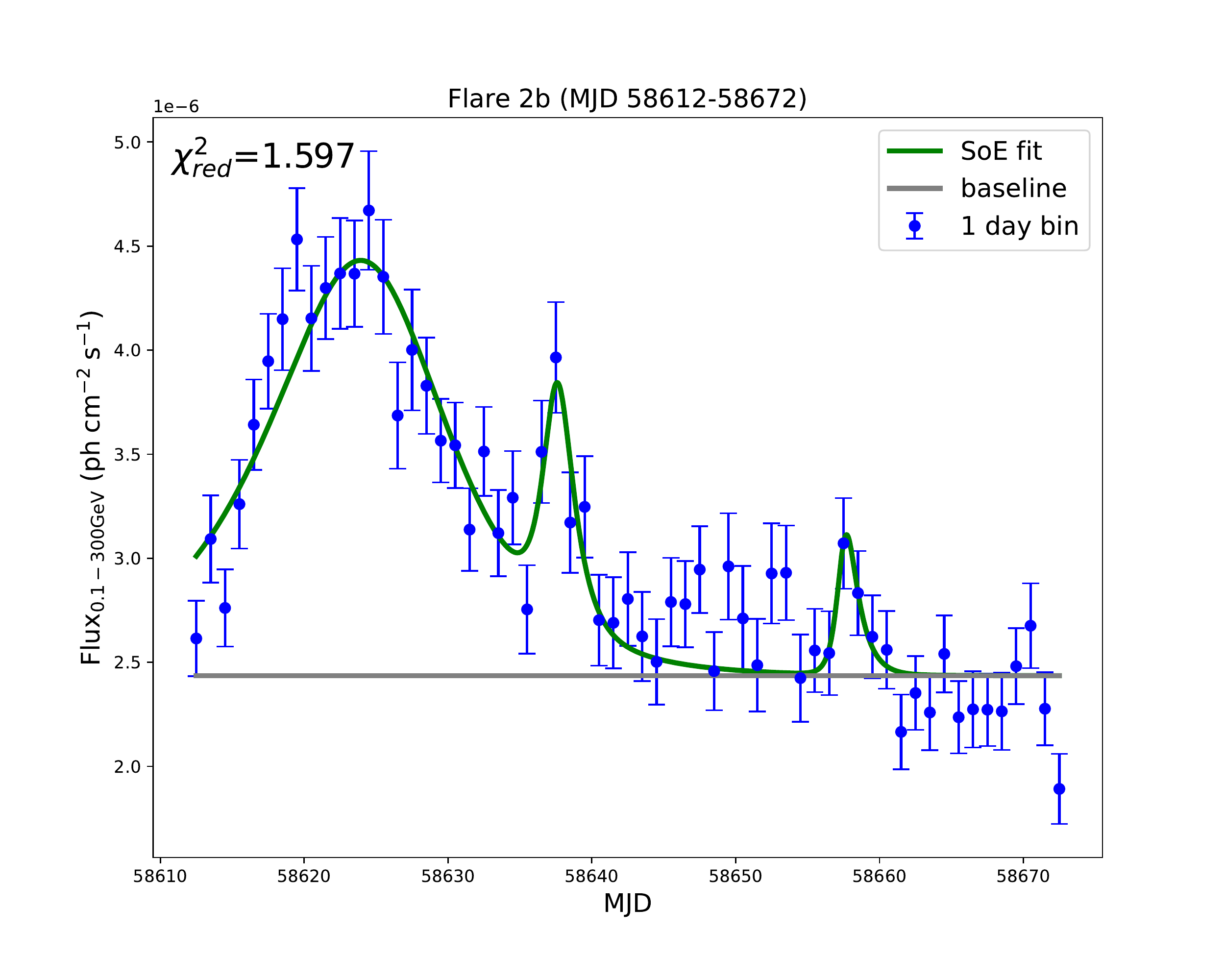}}\\[-5ex]
    \subfloat[]{\includegraphics[width =  \textwidth, height=0.46\textwidth ]{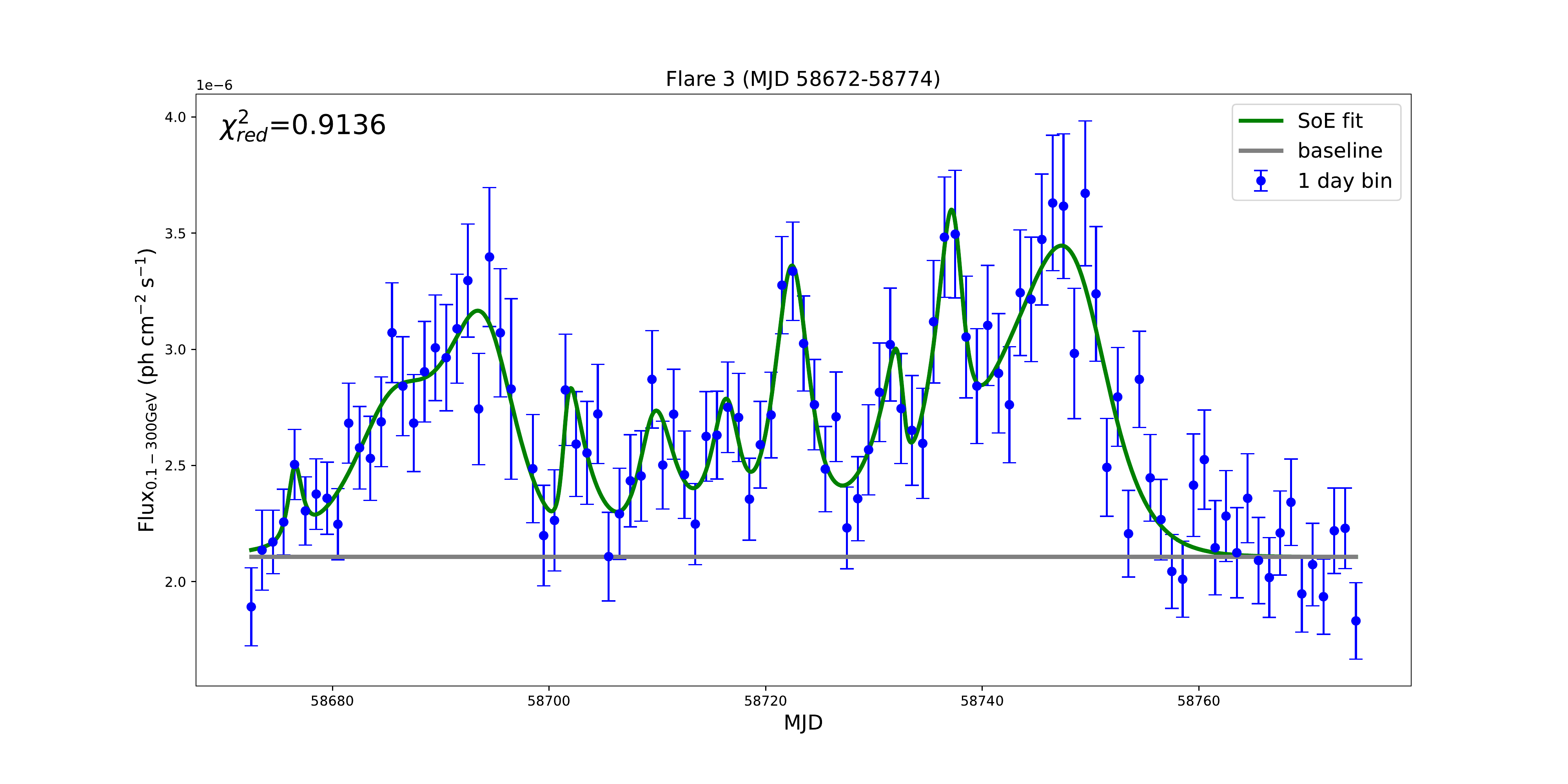}}

    \caption{The local peaks in the $\mathrm{\gamma-ray}$ light curve for each flare are fitted with the Sum of Exponentials (SoE) function. The individual Flares 1, 2, and 4 are subdivided into two parts (a and b) favor of a better fit. A total of 39 peaks were modeled and the corresponding rise and decay times were estimated. The reduced $\chi^2$ ($\chi^2_{red}$) values are calculated to estimate the goodness-of-fit and are mentioned in each plot. Fitting for Flare 1a was not satisfactory due to data discontinuities and is excluded from the analysis.}
    \label{fig_gamma_lc_fit}
\end{figure*}

\begin{figure*}\ContinuedFloat
    \centering
    \subfloat[]{\includegraphics[width = 0.5 \textwidth]{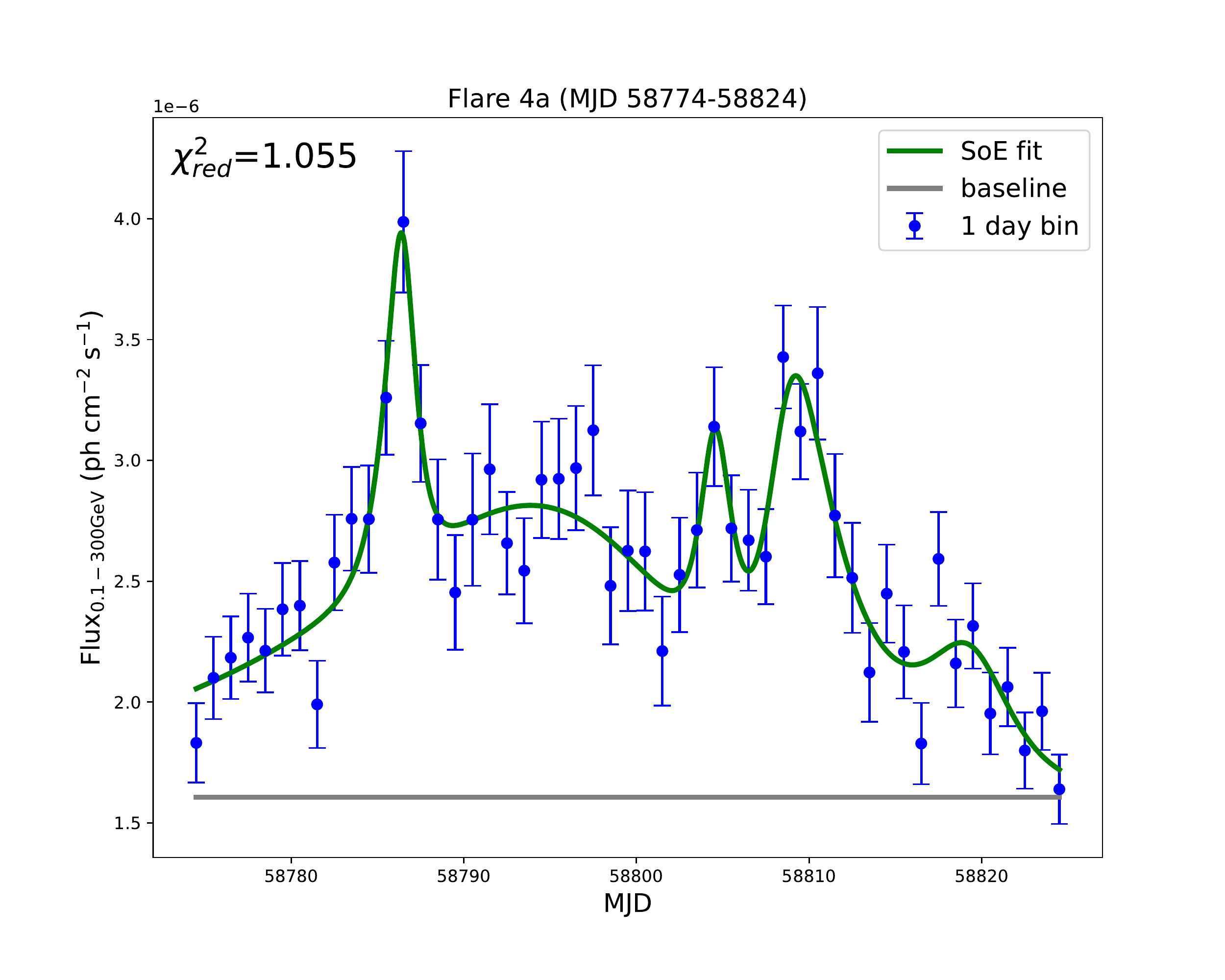}}
    \subfloat[]{\includegraphics[width = 0.5 \textwidth]{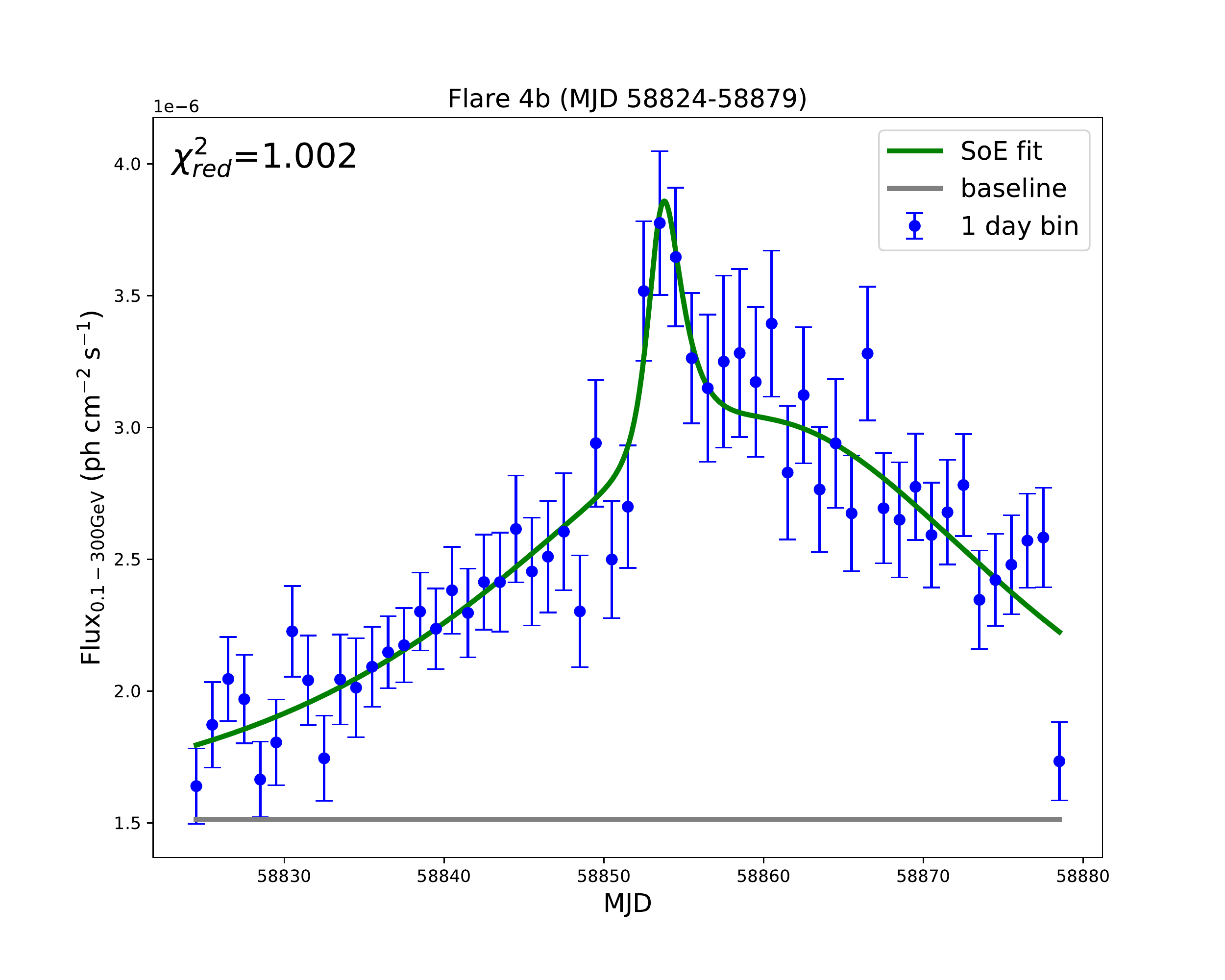}}\\[-5ex]
    \subfloat[]{\includegraphics[width =  \textwidth]{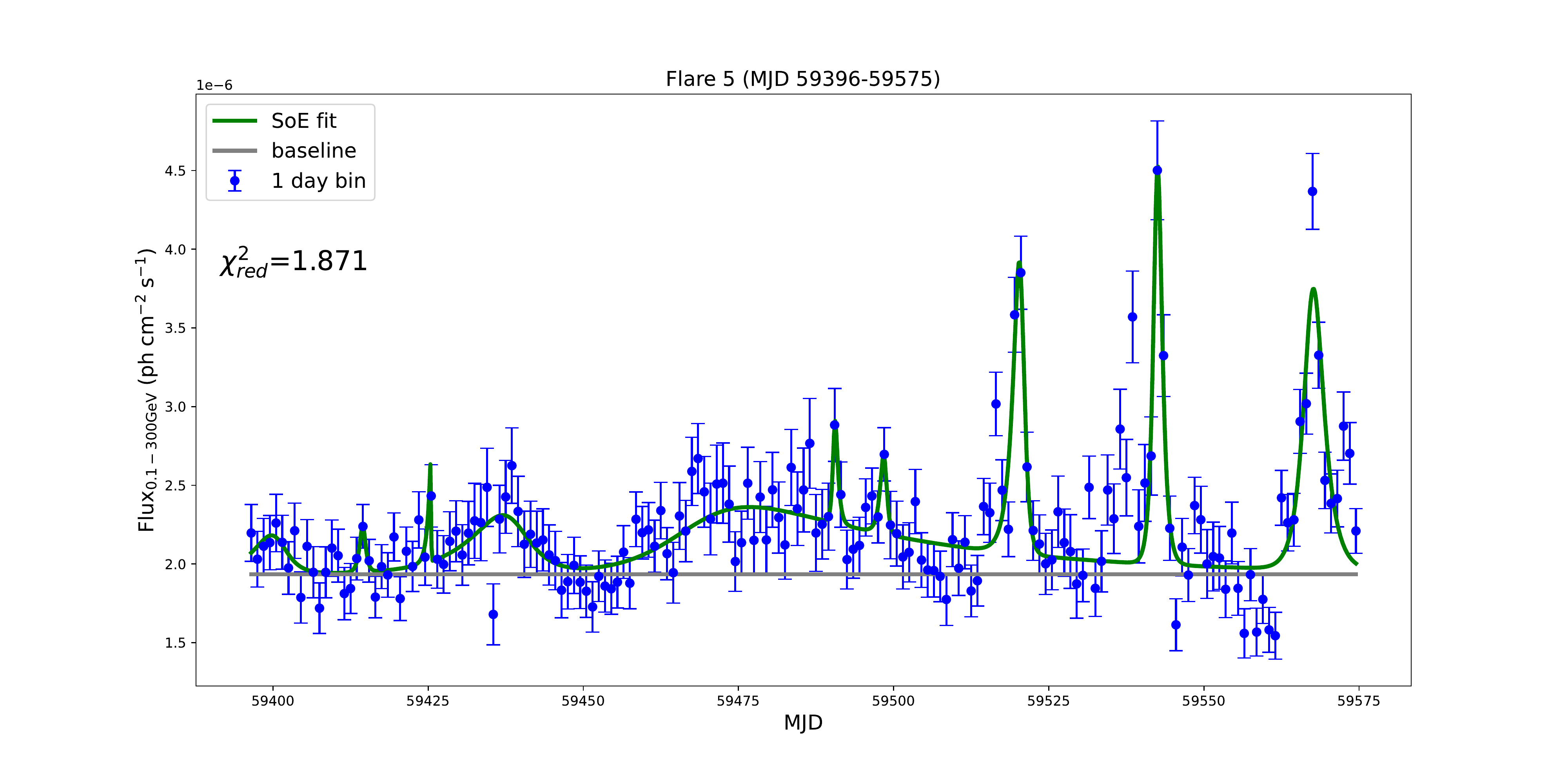}}
    \caption{(Continued)}
    \label{fig:my_label}
\end{figure*}

\begin{equation}
    A(t) = A_0 + \sum_{i=1}^{n} 2C_i \left( \exp\left(\frac{P_{ti}-t}{R_{ti}}\right) + \exp\left(\frac{t-P_{ti}}{D_{ti}}\right) \right)
\label{eqn:sum_of_exponetials}
\end{equation}
where $A(t)$ is the flux magnitude, $A_0$ is the continuum flux, $C_i$s are the peak constants, $P_{ti}$ is the peak time value while $R_{ti}$ and $D_{ti}$ are the rise and decay times respectively of $i$th peak in the period considered for fitting. A total of 39 peaks were modeled and the distribution of the obtained rise and decay times is shown in Figure \ref{fig:rise_and_decay_time_distribution}. The peaks were chosen such that the average of the five points with the current point at the center is the highest compared to immediate neighboring points. The distribution of both rise and decay times appears to be skewed towards the left with peaks around 2 days and some values extending up to 15 days.  The rise and decay time distribution is shown in Figure 2.

\subsection{Multi-wavelength variability}
\label{section:variability}
The variability in the light curve provides an indirect measurement of the size and location of the emission region where the broadband emission (light 
curves) are produced. 
The fast flux variability of the order of hours or minutes (\cite{Goyal_2017}, \cite{Goyal_2018}) scales suggest a very small region very close to the supermassive black hole (SMBH). Many models have been proposed to explain that. The best-accepted model is the shock-in-jet (\cite{1985ApJ...298..114M}) model where the shock can be produced at the base of the jet with local fluctuations. Another possible explanation has been proposed recently known as magnetic-reconnection which happens at farther distances down the jet (\cite{shukla2020gamma}).
The location of the emission region is another long-standing problem in blazar physics. In some sources, It has been found that under the one-zone leptonic scenario the emission region is generally located within the boundary of the broad-line region (BLR). However, there have been many studies that show the location of the emission region can also be far down the jet outside the BLR. 

The variability time scale can be estimated using the following expression,

\begin{equation}
    F_2 = F_1 2^{(t_2 - t_1)/t_d}
\end{equation}
where F$_2$ and F$_1$ are the fluxes measured at times t$_2$ and t$_1$ and the t$_d$ is the flux doubling/halving time. The minimum variability times are calculated for the wavelengths shown in Figure \ref{fig_broadband_lc} and the values are presented in Table \ref{tab:minimum_variability_times}. For the distribution of $t_d$ values see Appendix \ref{appendix_sec1}.

\begin{table}
\setlength{\extrarowheight}{4pt}
\begin{tabular}{ccc}
\hline
\textbf{Instrument}         & \textbf{Wavelength} & \textbf{Minimum variability time (days)} \\ \hline
Fermi-LAT                   & Gamma-ray          & $1.343 \pm 0.301$                        \\ \hline
Swift-XRT                   & X-ray              & $3.237 \pm 2.652$                        \\ \hline
\multirow{6}{*}{Swift-UVOT} & UVW2                & $0.156 \pm 0.068$                        \\
                            & UVM2                & $0.144 \pm 0.074$                        \\
                             & UVW1                & $0.105 \pm 0.098$                        \\
                            & U                   & $0.160 \pm 0.185$                        \\
                            & B                   & $0.131 \pm 0.058$                        \\
                            & V                   & $0.102 \pm 0.053$                        \\ \hline
\end{tabular}
\caption{\label{tab:minimum_variability_times}Minimum variability times from light curves of different wavelengths.}
\end{table}

\begin{table*}
\centering
{\scriptsize
\begin{center}
\renewcommand{\arraystretch}{1.5}
    \begin{tabular}{|c c c c c c c c|} 
    \hline
    \multicolumn{8}{|c|}{Power Law} \\ \hline 
    Flare & $\mathrm{F_{0.1-300GeV}}$ & Prefactor ($N_0$) & Index($\gamma$) & Scale ($E_0$) 
    & & TS & TS$_{curve}$\\
    & ($\mathrm{10^{-7}\:ph\:cm^{-2}\:s^{-1}}$) & ($\mathrm{10^{-10}\:ph\:cm^{-2}\:s^{-1}\: MeV^{-1}}$) &  & ($\mathrm{MeV}$) & &  & \\ \hline
    1 & 9.513 $\pm$ 0.008 & 2.022 $\pm$ 0.001 & -2.018 $\pm$ 0.0005 & \multirow{5}{*}{680.1} & & 14525.93& - \\
    2 & 11.085 $\pm$ 0.049 & 2.404 $\pm$ 0.009 & -1.996 $\pm$ 0.003 & & & 25411.73& - \\
    3 & 7.212 $\pm$ 0.046 & 1.501 $\pm$ 0.008 & -2.040 $\pm$ 0.004 & & & 13420.26 & -\\
    4 & 6.189 $\pm$ 0.053 & 1.335 $\pm$ 0.009 & -2.002 $\pm$ 0.006 & & & 11105.37& - \\ 
    5 & 3.300 $\pm$ 0.067 & 0.714 $\pm$ 0.012 & -2.001 $\pm$ 0.013 & & & 8909.44&- \\
    \hline
    
    \multicolumn{8}{|c|}{Broken Power Law} \\ \hline
        Flare &  $\mathrm{F_{0.1-300GeV}}$ & Prefactor ($N_0$) & Index1 ($\gamma_1$) & Index2 ($\gamma_2$) & Break Value ($E_b$) & TS & TS$_{curve}$ \\
         & ($\mathrm{10^{-7}\:ph\:cm^{-2}\:s^{-1}}$) & ($\mathrm{10^{-10}\:ph\:cm^{-2}\:s^{-1}\: MeV^{-1}}$) & &  & ($\mathrm{MeV}$) & &\\ \hline
         1 & 9.133 $\pm$ 0.022 & 0.3721 $\pm$ 0.0005 & -1.924 $\pm$ 0.001 & -2.234 $\pm$ 0.003 & 1692 $\pm$ 1.122 & 14447.65 &-156.56 \\
         2 & 10.287 $\pm$ 0.024 & 1.345 $\pm$ 0.002 & -1.821 $\pm$ 0.001 & -2.207 $\pm$ 0.002 & 998.3 $\pm$ 0.671 & 23525.53 &-3772.4 \\
         3 & 6.888 $\pm$ 0.036 & 1.352 $\pm$ 0.004 & -1.894 $\pm$ 0.003 & -2.181 $\pm$ 0.005 & 766.7 $\pm$ 1.143 & 13411.65 & -17.22\\
         4 & 5.564 $\pm$ 0.003 & 0.7686 $\pm$ 0.003 & -1.781 $\pm$ 0.0003 & -2.235 $\pm$ 0.0006 & 998.1 $\pm$ 0.1753 & 10659.25 & -892.24\\ 
         5 & 3.100 $\pm$ 0.085 & 0.390 $\pm$ 0.015 & -1.844 $\pm$ 0.026 & -2.189 $\pm$ 0.036 & 1005.0 $\pm$ 19.7 & 8931.38 & 43.88\\
         \hline
        
    \multicolumn{8}{|c|}{Log-Parabola} \\ \hline
        Flare &  $\mathrm{F_{0.1-300GeV}}$ & Norm ($N_0$) & $\alpha$ & $\beta$ & $E_b$ & TS & TS$_{curve}$ \\
         & ($\mathrm{10^{-7}\:ph\:cm^{-2}\:s^{-1}}$) & ($\mathrm{10^{-10}\:ph\:cm^{-2}\:s^{-1}\: MeV^{-1}}$) & &  & ($\mathrm{MeV}$) & &\\ \hline
         1 & 9.048 $\pm$ 0.056 & 2.185 $\pm$ 0.008 & 1.978 $\pm$ 0.003 & 0.5177 $\pm$ 0.0159 & \multirow{5}{*}{680.1} & 14461.04 & -129.78 \\
         2 & 10.450 $\pm$ 0.103 & 2.673 $\pm$ 0.019 & 1.949 $\pm$ 0.006 & 0.709 $\pm$ 0.0333 & & 23534.64& -3754.18 \\
         3 & 6.798 $\pm$ 0.014 & 1.647 $\pm$ 0.002 & 1.999 $\pm$ 0.001 & 0.6499 $\pm$ 0.0062 & & 13386.68& -67.16 \\
         4 & 5.269 $\pm$ 0.004 & 1.486 $\pm$ 0.001 & 1.887 $\pm$ 0.001 & -0.998 $\pm$ 0.002 & & 10425.39&-1359.96 \\ 
         5 & 2.900 $\pm$ 0.027 & 0.789 $\pm$ 0.005 & 1.912 $\pm$ 0.006 & 0.836 $\pm$ 0.028 & & 8560.39& -698.1 \\
         \hline
    
    \multicolumn{8}{|c|}{PLExpCutoff} \\ \hline
        Flare &  $\mathrm{F_{0.1-300GeV}}$ & Prefactor ($N_0$) & Index ($\gamma$) & Scale ($E_0$) & Cutoff ($E_c$) & TS & TS$_{curve}$ \\
         & ($\mathrm{10^{-7}\:ph\:cm^{-2}\:s^{-1}}$) & ($\mathrm{10^{-10}\:ph\:cm^{-2}\:s^{-1}\: MeV^{-1}}$) & &  & ($\mathrm{10^4\:MeV}$) && \\ \hline
         1 & 9.211 $\pm$ 0.208 & 2.154 $\pm$ 0.042 & -1.938 $\pm$ 0.018 & \multirow{5}{*}{680.1} & 2.992 $\pm$ 0.172 & 14486.28 & -79.3 \\
         2 & 10.688 $\pm$ 0.002 & 2.665 $\pm$ 0.0004 & -1.884 $\pm$ 0.0001 & & 1.860 $\pm$ 0.002 &  25454.22&84.98 \\
         3 & 6.905 $\pm$ 0.010 & 1.629 $\pm$ 0.002 & -1.937 $\pm$ 0.001 & & 2.054 $\pm$ 0.020 & 13347.70&-145.12 \\
         4 & 5.843 $\pm$ 0.006 & 1.467 $\pm$ 0.001 & -1.873 $\pm$ 0.001 & & 1.806 $\pm$ 0.009 & 11010.50&-189.74 \\ 
         5 & 3.100 $\pm$ 0.006 & 0.800 $\pm$ 0.001 & -1.852 $\pm$ 0.001 & & 1.444 $\pm$ 0.001 & 8904.12&-10.64 \\
         \hline
    \end{tabular}
    \caption{Spectral parameters of $\mathrm{\gamma-ray}$ SED fitted with four different models (Power Law, Broken Power Law, Log-Parabola, and PLExpCutoff). See Figure \ref{fig_gamma_sed} for the model plots.}
    \label{tab:gamma_ray_sed_param}
\end{center}
}
\end{table*}

\begin{figure*}
    \centering
    \includegraphics[width=0.96\textwidth]{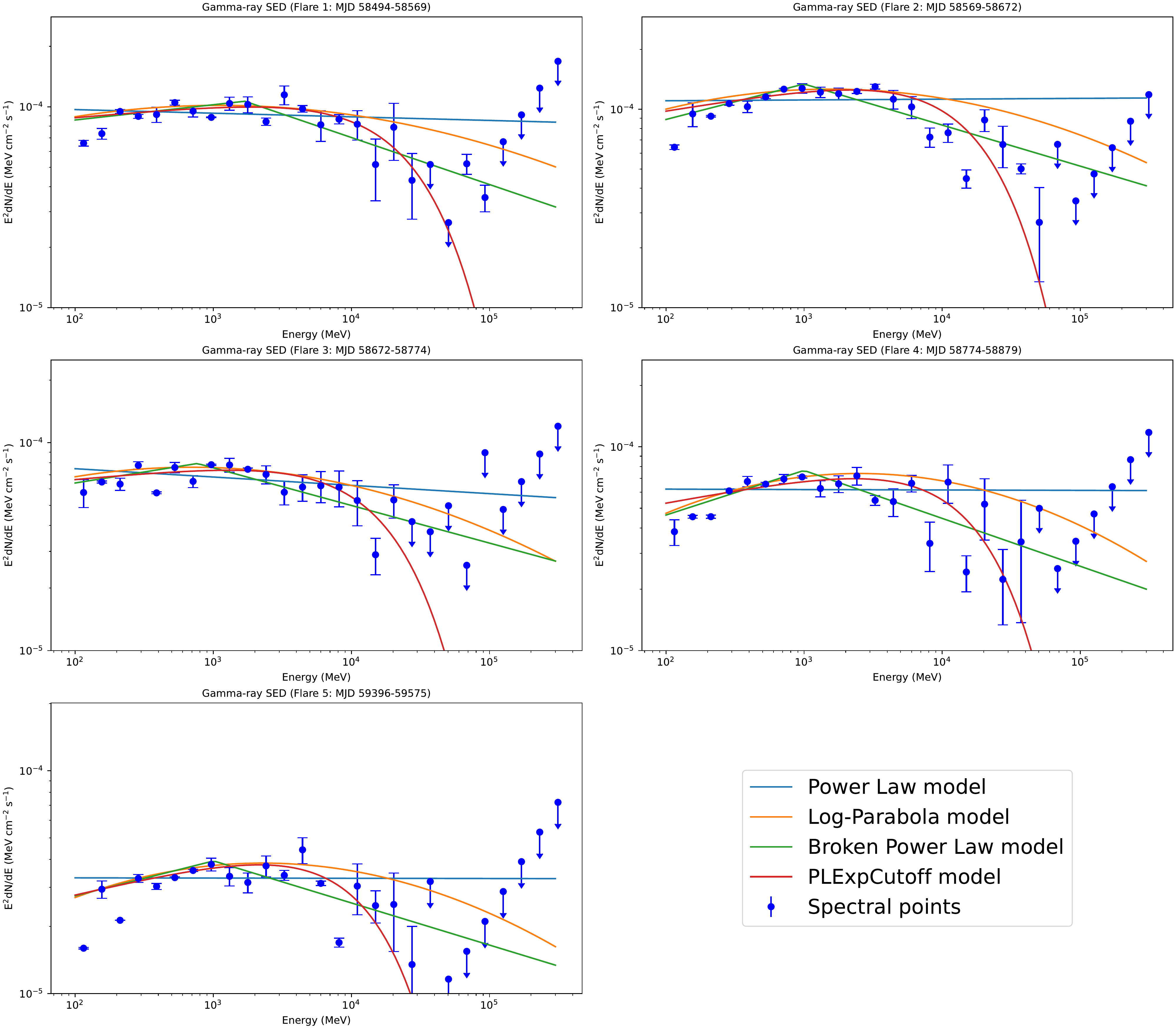}
    \caption{$\mathrm{\gamma-ray}$ SEDs obtained from Fermi-LAT data for each flare. The data is fitted with Power Law, Log-Parabola, Broken Power Law, and Power Law with Exponential Cutoff models. The 0.1-300 GeV energy range is divided into 28 bins using \href{https://fermi.gsfc.nasa.gov/ssc/data/analysis/user/likeSED.py}{likeSED.py} (a Fermi user-contributed tool) resulting in one spectral point for each bin.  The five plots in this figure have common x and y axes. The results of the fit are given in Table \ref{tab:gamma_ray_sed_param}. }
    \label{fig_gamma_sed}
\end{figure*}

\section{Gamma-ray Spectrum}\label{sec:gamma_ray_SED}
The $\gamma$-ray spectral analysis is done following the unbinned likelihood method. Analysis was performed using a user-contributed tool \textit{likeSED3.py}\footnote{\url{https://fermi.gsfc.nasa.gov/ssc/data/analysis/user/likeSED.py}} with the NewMinuit algorithm for the flares 1-5. During the modeling, the parameters of sources outside the 10-degree ROI were kept constant while the ones inside 10 degrees were allowed to vary. The spectrum of the source (4FGL J0348.5-2749) was then divided into 28 bins in the 0.1-300 GeV energy range giving the same number of spectral points. The Power Law (PL), Log-Parabola (LP), Broken Power Law (BPL), and Power Law with Exponential Cut-off (PLExpCutoff) models (see equations \ref{eqn:power_law}-\ref{eqn:plexpcutoff}) were used for fitting these points. The parameters for each of the fits are given in Table \ref{tab:gamma_ray_sed_param} and the corresponding model plots are shown in Figure \ref{fig_gamma_sed}.

\begin{enumerate}
    \item Power Law (PL):
    \begin{equation}
    \frac{dN}{dE} = N_0 (E/E_0)^{\gamma}
    \label{eqn:power_law}
\end{equation}
    \item Broken Power Law (BPL):
    \begin{equation}
\frac{dN}{dE} = N_0 \times
    \begin{cases}
    (E/E_b)^{\gamma_1},& \text{if } E < E_b\\
    (E/E_b)^{\gamma_2},              & \text{otherwise}
\end{cases}
\label{eqn:broken_power_law}
\end{equation}
    \item Log-Parabola (LP):
    \begin{equation}
    \frac{dN}{dE} = N_0 \left(\frac{E}{E_b}\right)^{-(\alpha + \beta \log{(E/E_b)})}
\label{eqn:log_parabola}
\end{equation}
    \item PLExpCutoff:
    \begin{equation}
    \frac{dN}{dE} = N_0 \left(\frac{E}{E_0}\right)^{\gamma_1} \exp{(-(E/E_c))} 
\label{eqn:plexpcutoff}
\end{equation}

The Test Statistic (TS) value was chosen as a measure for the quality of the model fit and is computed from the likelihood values obtained from the modeling. The Test Statistic (TS) is given by $\mathrm{TS = -2\ln{(L_{max,0}/L_{max,1})}}$, where $\mathrm{L_{max,0}}$ is the maximum likelihood value without a source at a position, and $\mathrm{L_{max,1}}$ is the maximum likelihood value with a source. We have also examined the curvature in the spectrum by estimating the TS$_{curve}$ = 2 [log $\mathcal{L}(M)$ - log $\mathcal{L}(PL)$], where $\mathcal{L}$ is the likelihood function (\citealt{2012ApJS..199...31N})  and $M$ represents the model used for fitting which can be LP, BPL or PLExpCutoff. A larger negative value suggests a better fit. Considering that, we noted that for flares 1 and 2, BPL gave a better fit, and flares 3, 4, and 5 are represented best by LP. Similar results were also reported for other FSRQs in \citet{Britto2016} and \citet{Prince2017}.

We also noticed that, in Table \ref{tab:gamma_ray_sed_param}, for the Power Law case when the source goes from a low flux state to a high flux state the spectral index becomes softer and softer, suggesting a softer-when-brighter behavior. Previously in 2018, \citet{angioni2019large} observed a harder-when-brighter trend during the large flaring state.
This is a rare behavior seen in FSRQ. Most of the FSRQ show a harder-when-brighter behavior as can be seen in  \citet{Abdo_2010, Britto2016} and \citet{Prince2017}. However, softer-when-brighter behavior is more common in BL Lac-type sources. Although, the distinction is not much clear as discussed in \citet{Giommi2021}.

\end{enumerate}
\section{Multi-frequency SED modelling }\label{sec4}

\begin{table*}
\setlength{\extrarowheight}{3pt}
\begin{tabular}{|cccccc|}
\hline
\multicolumn{1}{|c|}{\textbf{Symbol}}           & \multicolumn{1}{c|}{\textbf{Parameter {[}Units{]}}}                            & \multicolumn{1}{c|}{\textbf{Flare 1}} & \multicolumn{1}{c|}{\textbf{Flare 2}} & \multicolumn{1}{c|}{\textbf{Flare 3}} & \textbf{Flare 5} \\ \hline
\multicolumn{1}{|c|}{$\mathrm{\gamma_{min}}$}   & \multicolumn{1}{c|}{Low Energy Cut-Off {[}$10^1${]}}                           & \multicolumn{1}{c|}{1.77}             & \multicolumn{1}{c|}{5.42}             & \multicolumn{1}{c|}{3.45}             & 9.45             \\
\multicolumn{1}{|c|}{$\mathrm{\gamma_{max}}$}   & \multicolumn{1}{c|}{High Energy Cut-Off {[}$10^4${]}}                          & \multicolumn{1}{c|}{1.62}             & \multicolumn{1}{c|}{3.11}             & \multicolumn{1}{c|}{1.96}             & 5.81             \\
\multicolumn{1}{|c|}{N}                         & \multicolumn{1}{c|}{Electron Density {[}$\mathrm{10^4/cm^3}${]}}               & \multicolumn{1}{c|}{14.36}            & \multicolumn{1}{c|}{4.43}             & \multicolumn{1}{c|}{5.76}             & 9.46             \\
\multicolumn{1}{|c|}{$\mathrm{\gamma_{cut}}$}   & \multicolumn{1}{c|}{Turn-Over Energy {[}$10^3${]}}                             & \multicolumn{1}{c|}{8.02}             & \multicolumn{1}{c|}{4.26}             & \multicolumn{1}{c|}{3.12}             & 3.64             \\
\multicolumn{1}{|c|}{p}                         & \multicolumn{1}{c|}{Low Energy Spectral Slope}                                 & \multicolumn{1}{c|}{2.16}             & \multicolumn{1}{c|}{2.09}             & \multicolumn{1}{c|}{2.22}             & 2.14             \\
\multicolumn{1}{|c|}{$\mathrm{R'}$}             & \multicolumn{1}{c|}{Emission Region Size {[}$10^{15}$ cm{]}}                   & \multicolumn{1}{c|}{1.00}             & \multicolumn{1}{c|}{1.00}             & \multicolumn{1}{c|}{1.00}             & 0.37             \\
\multicolumn{1}{|c|}{$\mathrm{R_H}$*}           & \multicolumn{1}{c|}{Emission Region Position {[}$10^{17}$ cm{]}}               & \multicolumn{1}{c|}{1.00}             & \multicolumn{1}{c|}{1.00}             & \multicolumn{1}{c|}{1.00}             & 1.00             \\
\multicolumn{1}{|c|}{B}                         & \multicolumn{1}{c|}{Magnetic Field {[}gauss{]}}                                & \multicolumn{1}{c|}{0.56}             & \multicolumn{1}{c|}{0.77}             & \multicolumn{1}{c|}{0.92}             & 1.27             \\
\multicolumn{1}{|c|}{$\mathrm{\delta_D}$*}      & \multicolumn{1}{c|}{Doppler factor}                                            & \multicolumn{1}{c|}{60.00}            & \multicolumn{1}{c|}{60.00}            & \multicolumn{1}{c|}{60.00}            & 60.00            \\
\multicolumn{1}{|c|}{$\mathrm{z_{cosm}}$*}      & \multicolumn{1}{c|}{Redshift}                                                  & \multicolumn{1}{c|}{0.991}            & \multicolumn{1}{c|}{0.991}            & \multicolumn{1}{c|}{0.991}            & 0.991            \\
\multicolumn{1}{|c|}{$\mathrm{\tau_{BLR}}$}     & \multicolumn{1}{c|}{Optical depth}                                             & \multicolumn{1}{c|}{0.07}             & \multicolumn{1}{c|}{0.07}             & \multicolumn{1}{c|}{0.07}             & 0.07             \\
\multicolumn{1}{|c|}{$\mathrm{R_{BLR_{in}}}$*}  & \multicolumn{1}{c|}{Inner radius of BLR {[}$10^{17}$ cm{]}}                    & \multicolumn{1}{c|}{1.18}             & \multicolumn{1}{c|}{1.18}             & \multicolumn{1}{c|}{1.18}             & 1.18             \\
\multicolumn{1}{|c|}{$\mathrm{R_{BLR_{out}}}$*} & \multicolumn{1}{c|}{Outer radius of BLR {[}$10^{18}$ cm{]}}                    & \multicolumn{1}{c|}{1.18}             & \multicolumn{1}{c|}{1.18}             & \multicolumn{1}{c|}{1.18}             & 1.18             \\
\multicolumn{1}{|c|}{$\mathrm{L_{Disk}}$*}      & \multicolumn{1}{c|}{Disk luminosity {[}$10^{45}$ erg/s{]}}                     & \multicolumn{1}{c|}{1.40}             & \multicolumn{1}{c|}{1.40}             & \multicolumn{1}{c|}{1.40}             & 1.40             \\
\multicolumn{1}{|c|}{$\mathrm{T_{Disk}}$}       & \multicolumn{1}{c|}{Disk temperature {[}$10^5$ K{]}}                           & \multicolumn{1}{c|}{10.00}            & \multicolumn{1}{c|}{10.00}            & \multicolumn{1}{c|}{9.78}             & 6.12             \\ \hline
\multicolumn{6}{|c|}{\textbf{Energy Densities}}                                                                                                                                                                                                                             \\ \hline
\multicolumn{1}{|c|}{\textbf{Symbol}}           & \multicolumn{1}{c|}{\textbf{Parameter {[}Units{]}}}                            & \multicolumn{1}{c|}{\textbf{Flare 1}} & \multicolumn{1}{c|}{\textbf{Flare 2}} & \multicolumn{1}{c|}{\textbf{Flare 3}} & \textbf{Flare 5} \\ \hline
\multicolumn{1}{|c|}{$\mathrm{U'_{BLR}}$}       & \multicolumn{1}{c|}{BLR Energy Density {[}$\mathrm{erg/cm^3}${]}}      & \multicolumn{1}{c|}{14.86}            & \multicolumn{1}{c|}{14.86}            & \multicolumn{1}{c|}{14.86}            & 14.86            \\
\multicolumn{1}{|c|}{$\mathrm{U'_{Disk}}$}      & \multicolumn{1}{c|}{Disk Energy Density {[}$\mathrm{erg/cm^3}${]}}     & \multicolumn{1}{c|}{5.79}             & \multicolumn{1}{c|}{5.79}             & \multicolumn{1}{c|}{5.79}             & 5.79             \\
\multicolumn{1}{|c|}{$\mathrm{U'_e}$}           & \multicolumn{1}{c|}{Electron Energy Density {[}$\mathrm{erg/cm^3}${]}} & \multicolumn{1}{c|}{8.77}             & \multicolumn{1}{c|}{7.17}             & \multicolumn{1}{c|}{5.28}             & 22.38            \\
\multicolumn{1}{|c|}{$\mathrm{U'_B}$}           & \multicolumn{1}{c|}{Magnetic Field Energy Density {[}$10^{-2}\: \mathrm{erg/cm^3}${]}}    & \multicolumn{1}{c|}{1.24}             & \multicolumn{1}{c|}{2.37}             & \multicolumn{1}{c|}{3.37}             & 6.37             \\ \hline
\multicolumn{6}{|c|}{\textbf{Jct Power}}                                                                                                                                                                                                                                    \\ \hline
\multicolumn{1}{|c|}{\textbf{Symbol}}           & \multicolumn{1}{c|}{\textbf{Parameter {[}Units{]}}}                            & \multicolumn{1}{c|}{\textbf{Flare 1}} & \multicolumn{1}{c|}{\textbf{Flare 2}} & \multicolumn{1}{c|}{\textbf{Flare 3}} & \textbf{Flare 5} \\ \hline
\multicolumn{1}{|c|}{$\mathrm{P_{jet}}$}        & \multicolumn{1}{c|}{Total Jet Power {[}$10^{45}$ ergs/s{]}}                    & \multicolumn{1}{c|}{2.98}             & \multicolumn{1}{c|}{2.44}             & \multicolumn{1}{c|}{1.80}             & 1.05             \\
\multicolumn{1}{|c|}{$\mathrm{P_e}$}            & \multicolumn{1}{c|}{Jet Power in Electrons {[}$10^{45}$ ergs/s{]}}             & \multicolumn{1}{c|}{2.97}             & \multicolumn{1}{c|}{2.43}             & \multicolumn{1}{c|}{1.79}             & 1.05             \\
\multicolumn{1}{|c|}{$\mathrm{P_B}$}            & \multicolumn{1}{c|}{Jet Power in Magnetic Field {[}$10^{43}$ ergs/s{]}}        & \multicolumn{1}{c|}{0.42}             & \multicolumn{1}{c|}{0.80}             & \multicolumn{1}{c|}{1.14}             & 0.30             \\ \hline
\end{tabular}
\caption{\label{tab:SED-model-parameters} The parameters obtained from modeling the multi-frequency SEDs of Flares 1, 2, 3, and 5 using JetSeT. The Power Law with Cutoff (PLC) model was chosen for the radiating electrons. The energy densities and jet powers due to electrons and emission region magnetic field are subsequently computed. Additionally, the energy density of the Broad Line Region (BLR) and the accretion disk is computed in the rest frame of the emission region. In the top section, parameters with (*) are fixed during modeling.}
\end{table*}

\begin{figure*}
    \centering
    \subfloat[]{\includegraphics[width = 0.5 \textwidth]{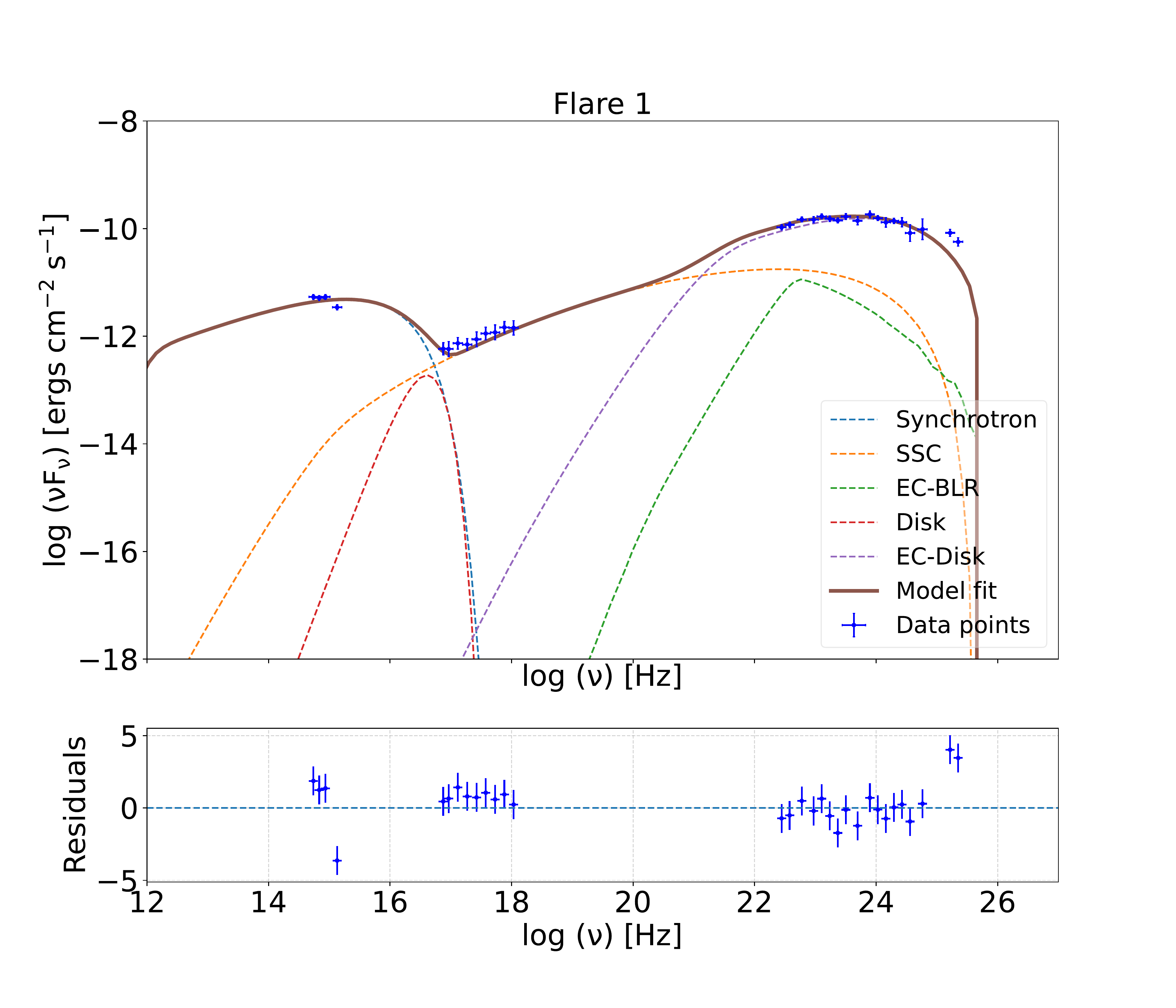}}
    \subfloat[]{\includegraphics[width = 0.5 \textwidth]{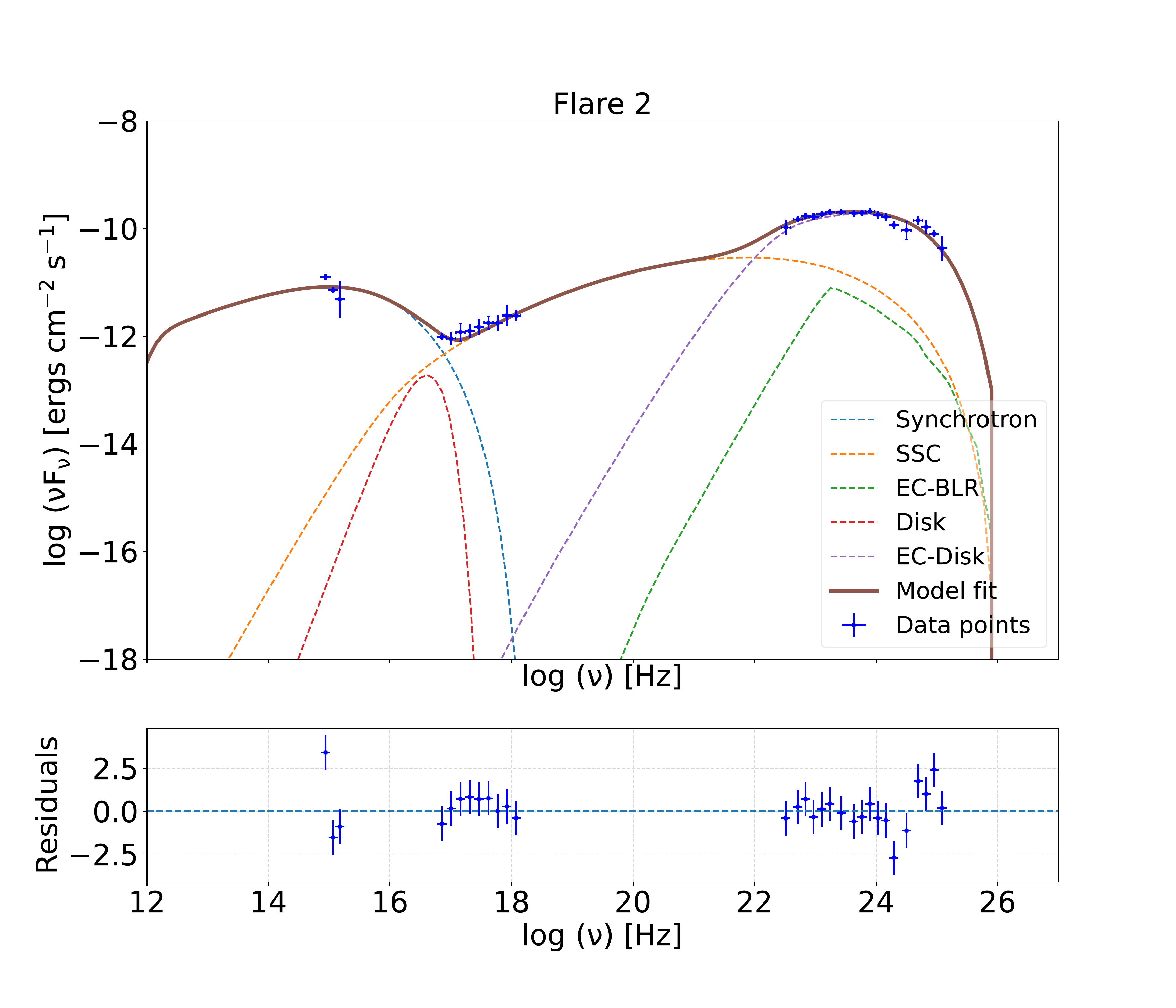}}\\   \subfloat[]{\includegraphics[width = 0.5 \textwidth]{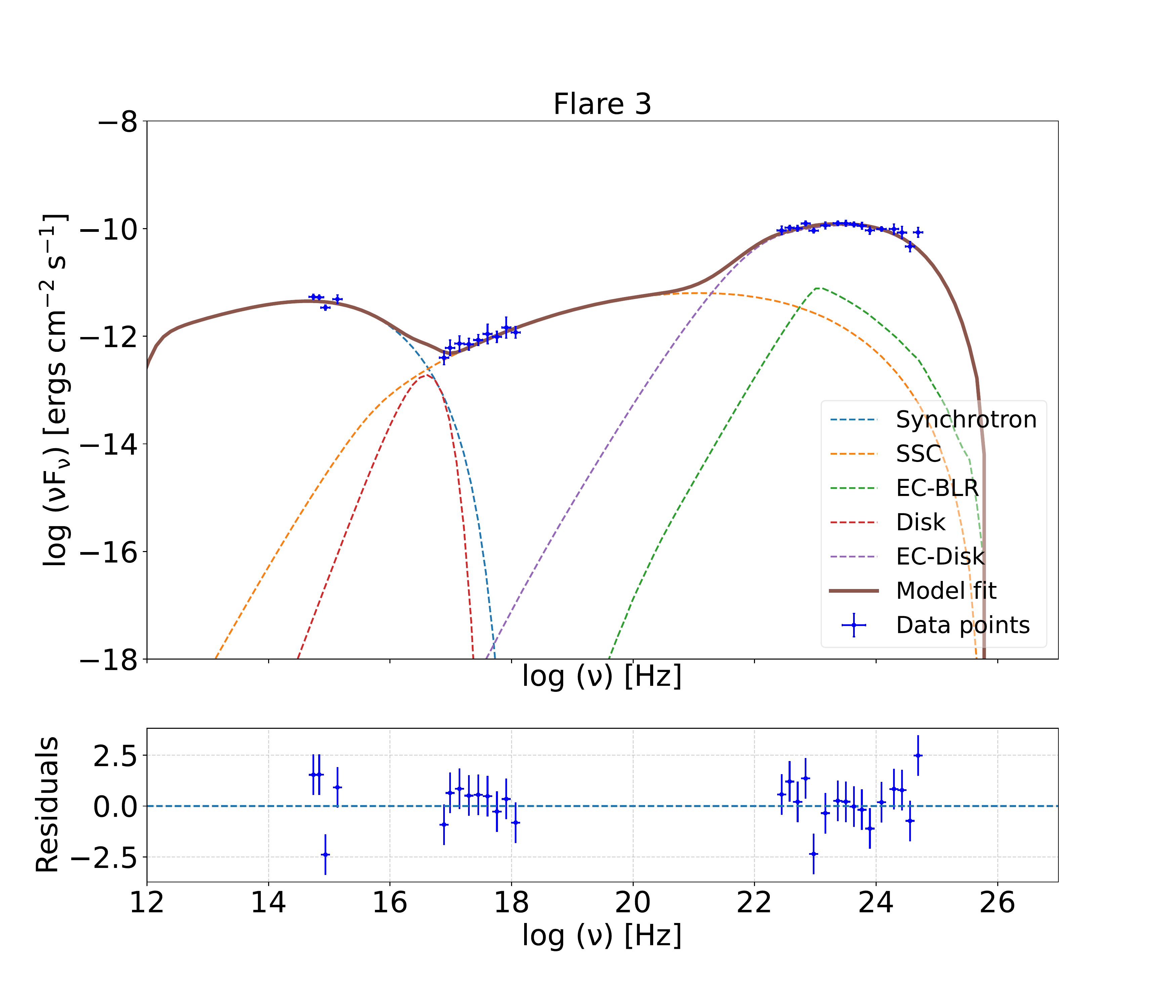}}
    \subfloat[]{\includegraphics[width = 0.5 \textwidth]{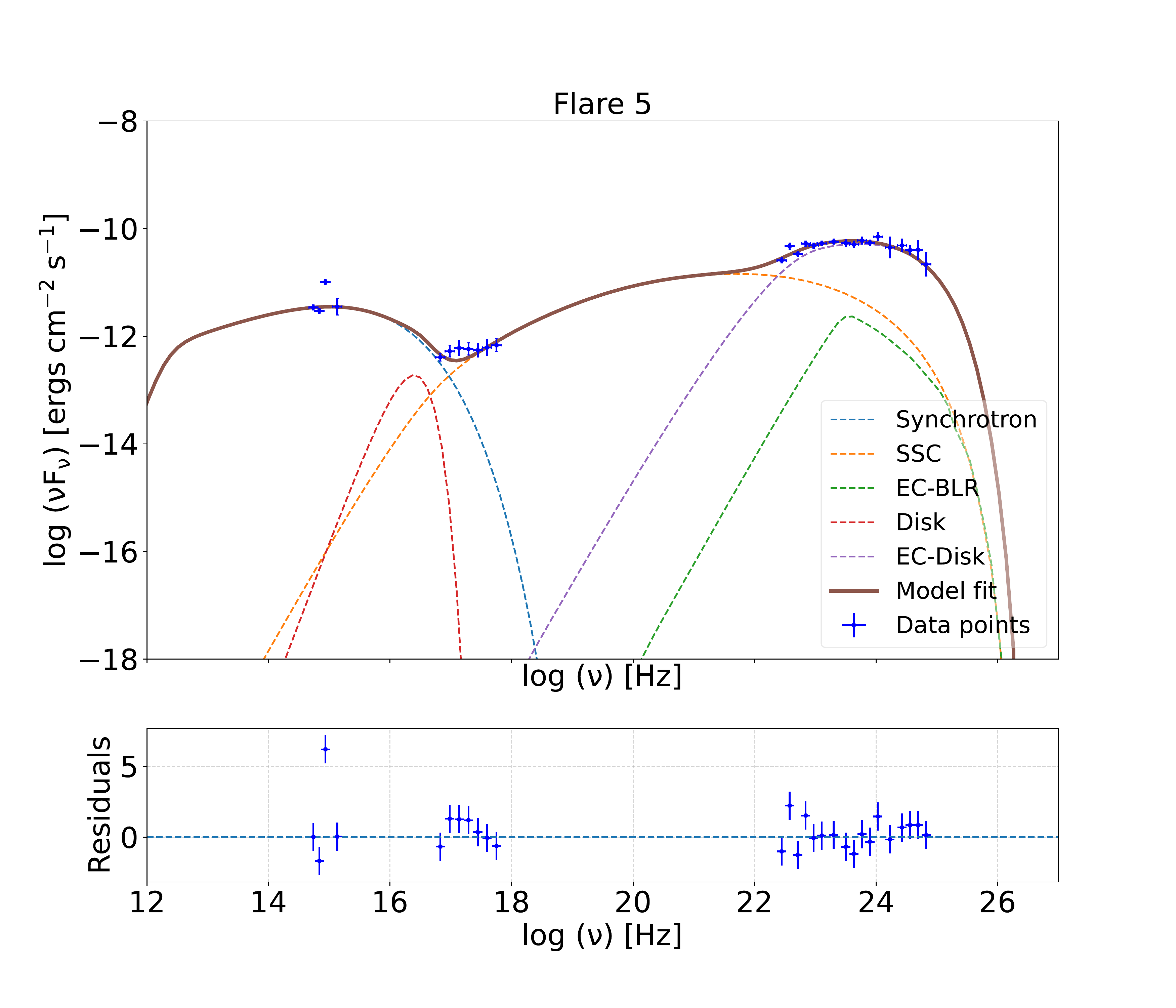}}
    
    \caption{Multi-frequency SEDs obtained from Fermi-LAT, Swift-XRT, and Swift-UVOT for Flares 1, 2, 3, and 5 are modeled using \href{https://jetset.readthedocs.io/en/1.1.2/}{JetSeT}.  The three clusters of data (from left to right) are from Swift-UVOT, Swift-XRT, and Fermi-LAT respectively.}
    \label{fig_broadband_sed_fit}
\end{figure*}

The emission mechanisms of a blazar can be understood better from the modeling of broadband spectral energy distribution. We obtained the broadband SED using data from Fermi-LAT, Swift-XRT, and Swift-UVOT for each of the flare periods shown in Figure \ref{fig_broadband_lc} except Flare 4 due to the lack of Swift data in that period. A typical broadband SED of a blazar has a two-hump structure with the lower energy hump associated with Synchrotron emission and the Inverse Compton scattered photons are assumed to be responsible for the higher energy hump.  The seed photons for the IC scattering could be from inside the jet (i.e. synchrotron photon) or outside the jet from BLR or DT, and Accretion disk.

We used JetSeT\footnote{\url{https://andreatramacere.github.io/jetsetdoc/html/index.html}} (\cite{2020ascl.soft09001T}, \cite{2011ApJ...739...66T}, \cite{2009A&A...501..879T}) code for modelling the broadband SED. The Power Law with Cut-off (PLC) model was considered for the electron distribution with low energy spectral slope $p$. The non-thermal Synchrotron radiation is assumed to be responsible for the lower energy emissions i.e., the first hump in the SED. The higher energy photons are produced via the Synchrotron Self-Compton (SSC) and External Compton (EC) emission mechanisms. These two mechanisms are special cases of the Inverse Compton mechanism and are assumed to be responsible for the second hump. The seed photons for SSC emissions are the Synchrotron photons which are produced when electrons and positrons are accelerated in the strong magnetic field in the emission region near the central black hole. For EC emissions, the photon inputs are mainly sourced from (1) the direct emission from the disk, (2) the reprocessed emission from BLR in the optical-UV frequency range, and (3) the reprocessed emission in the infrared region from the dusty torus. JetSeT implements the modeling of BLR using the method in \cite{2003APh....18..377D} and the disk emissions are modeled using chapter 5 of \cite{frank2002accretion} (for more detail see the description of JetSet).

The model considers a spherically symmetric region with a radius $\mathrm{R'}$ and  electron density $\mathrm{N}$ as the emission region. It has a speed $\beta c$ with respect to the observer, an entangled magnetic field B, and a bulk Lorentz factor $\Gamma = 1/\sqrt{1-\beta^2}$. The region moves at a small angle $\theta$ with the observer's line of sight and is located at a distance $\mathrm{R_H}$ from the central black hole. The corresponding Doppler factor is $\mathrm{\delta_D = 1/(\Gamma(1-\beta \cos{\theta}))}$.  In our case, we use $\Gamma = 60$ and take $\Gamma = \delta_D$ (\cite{angioni2019large}) which is quite high for an FSRQ type blazar (\citealt{2009A&A...494..527H, Wu_2018}). The synchrotron emission depends on the magnetic field and the speed of the relativistic electrons and is bounded by the Lorentz factors $\mathrm{\gamma_{min}}$ and $\mathrm{\gamma_{max}}$.  The Broad Line Region (BLR) has inner and outer radii $\mathrm{(R_{BLR_{in}}}$ \& $\mathrm{{R_{BLR_{out}})}}$ and an optical depth of $\mathrm{\tau_{BLR}}$. The primed quantities are in the comoving frame of the emission region. The emission region in the model is close to the disk and is assumed to be well inside the BLR (see Table \ref{tab:SED-model-parameters}) resulting in a relatively large contribution of Compton scattering from the accretion disk compared to the BLR and the dust torus. The accretion disk has a luminosity $\mathrm{L_{Disk}}$ and in a temperature $\mathrm{T_{Disk}}$. The Synchrotron emission was restricted to the frequency range $10^{10}-10^{18}$ Hz for effective fitting. Similarly, we constrained the Inverse Compton frequency range to $10^{21}-10^{29}$ Hz. In the case of FSRQs, generally, EC is the dominant mechanism for $\gamma$-ray emission, as is evident for this source also. The seed photons for inverse Compton scattering are provided by the accretion disk as shown in Figure \ref{fig_broadband_sed_fit}.

The energy densities in the accretion disk ($\mathrm{U'_{Disk}}$) and the BLR ($\mathrm{U'_{BLR}}$) are calculated directly by the model using JetSeT.  The ratio $\mathrm{U'_{Disk}/U'_{BLR}}$ is less than one for all the Flares (see Table \ref{tab:SED-model-parameters}). As expected the lower frequency emissions are dominated by Synchrotron and SSC components. The total power of the jet is estimated from the energy densities of leptons ($U'_e$), cold protons ($U'_p$), and magnetic field ($U'_B$) in the co-moving frame of the jet using the relation

\begin{equation}
    P_{jet} = \pi R^2 c^2 \Gamma^2 (U'_e + U'_p + U'_B)
\label{eqn:jet_power}
\end{equation}

while the powers due to magnetic field ($P_B$) and leptons ($P_e$) are given by 
\begin{equation}
    P_B = \frac{1}{8} \;c \; R^2 \; \Gamma^2 \; B^2
\end{equation}
\begin{equation}
    P_e = \frac{3c\Gamma^2}{4 R} \int_{E_{min}}^{E_{max}} E\; PLC(E) dE
\end{equation}

where $PLC(E)$ is the Power Law with a Cutoff model as a function of energy and $(E_{min}, E_{max})$ represents the energy range of the leptons considered for modeling. The energy densities and powers differ by a factor $\pi R^2 c^2 \Gamma^2$ (see equation \ref{eqn:jet_power}). For example, $P_B$ and $U'_B$ have the relation, $P_B = \pi R^2 c^2 \Gamma^2 U'_B$. The energy densities $\mathrm{U'_{e}, U'_{p}}$, and $\mathrm{U'_{B}}$ are also estimated by the model and the contribution from $\mathrm{U'_{p}}$ is not included in this model. Consequently, the relation between the jet powers is $\mathrm{P_{jet} \approx P_{e} + P_{B}}$. The respective values of parameters, energy densities, and jet power are mentioned in Table \ref{tab:SED-model-parameters}. The modeling was performed for different values of $\mathrm{\gamma_{min}}$ and $\mathrm{\gamma_{max}}$ and are adjusted for the best fit. The inner radius of the BLR region is set to $1.18 \times 10^{17} \: cm$ using the scaling relation $R_{BLR} \sim (L_{Disk}/10^{45})^{1/2} \: cm$ (\cite{Ghisellini2009Aug}) and $L_{Disk} = 1.4 \times 10^{45} \: erg/s$ (\cite{angioni2019large}). The $L_{Disk}$ value is fixed in this model. $R_{BLR_{out}}$ is set to an order of magnitude higher than $R_{BLR_{in}}$ and $R_H$ is set to the default value in JetSeT. The $P_{jet}$ value decreases with each flare but remains in the same order. The $\gamma_{cut}$ value remains of the same order for the entire period of analysis indicating similar cut-off energy. In the low energy region, the spectral slope ($p$) fluctuates minimally in the 2.09-2.22 range which can be attributed to the stability of seed photon emissions. Swift-XRT and Swift-UVOT data for the period in Flare 4 are not available for broadband SED modeling.

The variability time scales from section \ref{section:variability} can also be used to estimate the size of the emission region using $\mathrm{R' \le ct_{var}\delta_D/(1+z)}$  (\cite{Abdo_2011}). Using the minimum $\mathrm{\gamma-ray}$ variability ($t_{var} = 1.343 \; days$) we get a value of $1.05 \times 10^{17} \; cm$ for $R'$ and the corresponding values from the modeling are in the range $0.37-1.00 \times 10^{15} \; cm$ similar to the values in \cite{angioni2019large} where $R'$ is described as the comoving radius of the blob. Performing similar calculations using the values from Table \ref{tab:minimum_variability_times}, we find the values of $R'$ to be in the range $0.08-2.53 \times 10^{17} \; cm$ in some cases an order of magnitude higher than the modeling results this could possibly happen due to lack of well-sampled data. 

The distance of the emission region from the central black hole can be estimated from $\mathrm{R_H < c \Gamma^2 t_{var}/(1+z)}$ (\cite{Abdo_2011}). Taking $t_{var} = 1.343 \; days$ ($\gamma-ray$ minimum variability), the estimated upper limit of $\mathrm{R_H}$ value is $\mathrm{6.29 \times 10^{18} \; cm}$ and $R_H = 1.0 \times 10^{17} \; cm$ from the modelling. For the rest of the wavelengths in Table \ref{tab:minimum_variability_times}, the value range is $4.8-7.5 \times 10^{17} \; cm$. Here we noticed that for though gamma-ray the location is not exactly matching with the modeling result but the value estimated with all other emissions is consistent with the modeling result. We believed that this could happen because of many free parameters in the SED modeling. During the modeling, we noticed that the disk photon contributes more than the BLR photons and the gamma-ray is explained by EC-disk. This could be possible when the emission blob is too close to the central source. In our result, we see that the location of the blob is closer than the BLR location ($>$10$^{17}$) measured from the central source.

\section{Power Spectral Density}\label{sec5}
The variability in the source light curve can also be quantified by the power spectral density or PSD, which determines the amplitude of variation in the temporal light curve as a function of Fourier frequency or variability time scales (\citealt{2019ApJ...885...12R}). PSD is important to understand the average properties of the variability, whereas the source light curve could be thought of as only a single realization of an underlying stochastic process as shown by \citet{2003MNRAS.345.1271V}.  
We have derived the power spectral density or PSD using the discrete Fourier transformation (DFT). The shape of the PSD is best fitted with power law using the "power spectrum response" (PSRESP)" method. 

For an evenly sampled light curve f(t$_{i}$) the RMS-normalized PSD is the squared modulus of the DFT. Assuming the light curve is sampled over time steps t$_{i}$ and has a total number of points N, the RMS-normalized PSD can be defined as,

\begin{equation}
    P(\nu{_k}) = \frac{2 T}{\mu^2 N^2} \Bigg\{ \Big[ \sum_{i=1}^{N} f(t_{i}) cos(2\pi\nu_{k}t_i)     \Big]^2 + \Big[ \sum_{i=1}^{N} f(t_{i}) sin(2\pi\nu_{k}t_i)     \Big]^2  \Bigg \}
\end{equation}
where $\mu$ is the mean of the light curve and $\nu_k$ = $i$/T, $i$ = 1, 2, 3,....N/2 with the maximum frequency, Nyquist frequency, $\nu_{Nyq}$ = N/2T. 

The constant noise level is also estimated in form of normalized Poisson noise using the relation,

\begin{equation}
    P_{Poisson} = \frac{\sigma^2_{err}}{\mu^2(\nu_{Nyq} - \nu_{min})}
\end{equation}
where $\sigma^2_{err}$ is the mean-variance of the measurement uncertainty.

Further, we have used the "Power Spectral Response (PSRESP)" method to tackle the distortions in PSD caused by the Fourier transform and estimate the best fit power law slope to the PSD. This method has already been used by many authors and is currently one of the best methods to describe the best fit PSD (\citealt{Uttley2002, Chatterjee2008, Max2014, Meyer_2019, Bhattacharyya_2020, Goyal_2022}). The detailed procedure of the PSRESP method can be seen in \citep{Bhattacharyya_2020, Goyal_2022}. In the PSRESP method, we choose the range of PSD slope starting from 0.5 to 3.0 with steps 0.05 and corresponding to each slope success fraction is defined (see \citet{Bhattacharyya_2020}  for more details on success fraction). The best fit PSD slope is determined as 2.15$\pm$0.87. As it has been seen that the blazar variability is a stochastic process and can be fitted with a single Power Law. The PSD slope, $\beta$ =1 corresponds to pink or flicker noise, and $\beta$ = 2  represents the red noise. In our case, $\beta$ covers the range 1.28$-$3.02 suggesting variability in the blazar is a stochastic process of correlated colored-noise type (\citealt{Goyal2020}).
The PSD can be used to find out the characteristic time-scale in the system and that can be used to constrain the size of the emission region. The characteristic time scale can be derived when the PSD deviates from the Power Law shape and show some kind of break in the power spectrum. Generally, the breaking time scale can be characterized as the time scale of variability in the source or the particle cooling or escape time scales (\citealt{2011A&A...531A.123K, 2014ApJ...786..143S, 2014ApJ...791...21F, 2016MNRAS.458.3260C, 2017ApJ...849..138K, 2018ApJ...859L..21C, 2019ApJ...885...12R, Bhattacharyya_2020}). It suggests that even longer data sets are required to see any break in the PSD and hence the characteristic time scale. 

\begin{figure}
    \centering
    \includegraphics[width=0.5\textwidth]{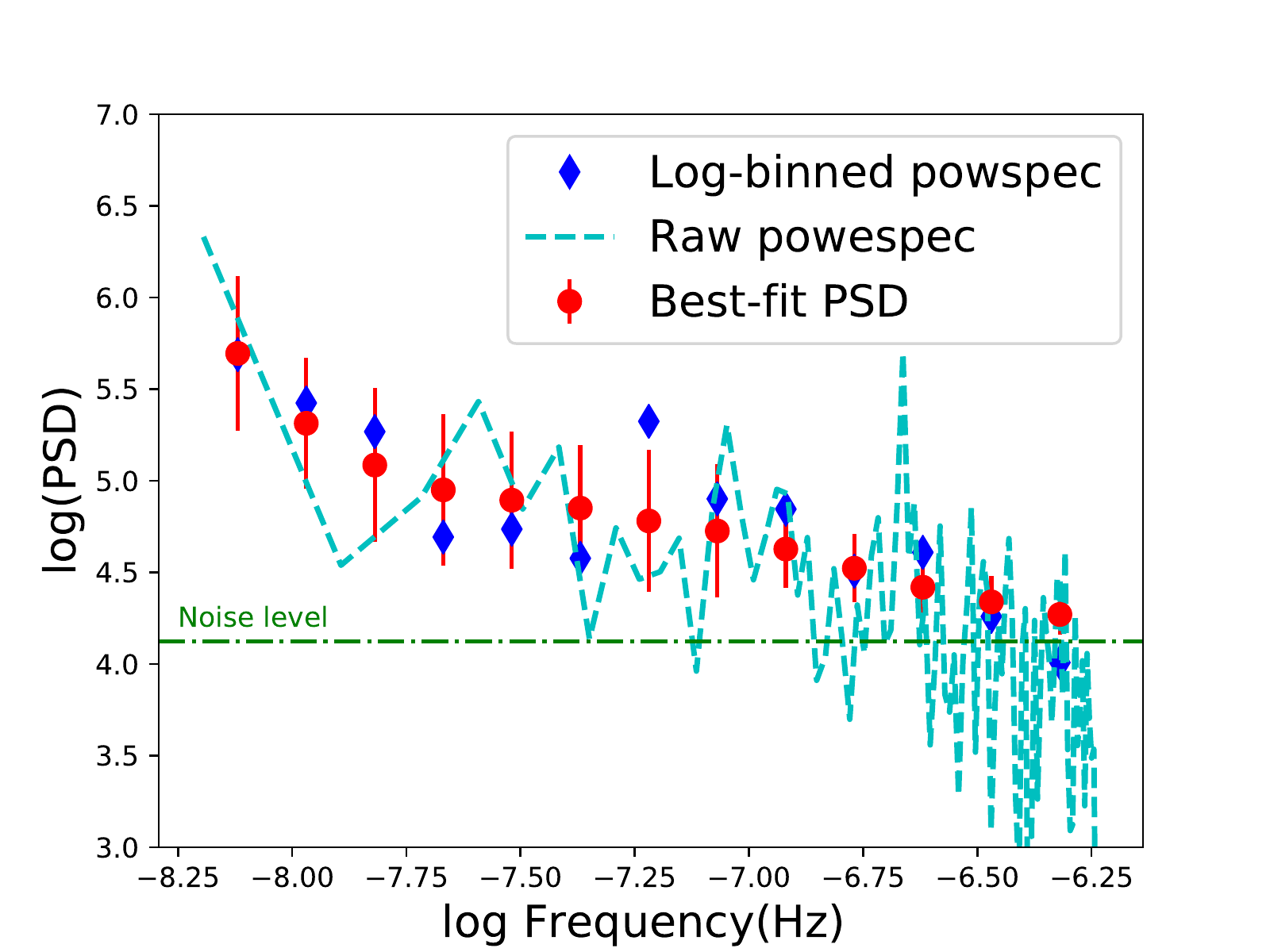}
    \caption{Best it PSD for the gamma-ray light curve derived using the PSRESP method.}
    \label{fig:PSD}
\end{figure}

The PSD is derived for the long-term gamma-ray light curve and the resultant PSD is shown in Figure \ref{fig:PSD}. 
The PSD can also be used to characterize the variability in other wavebands. The earlier results suggest that the variability in high energy bands (X-ray and gamma-ray) is characterized by pink or flicker noise (\citealt{Abdo_2010, Isobe_2014, Abdollahi_2017} ) and in lower energy (radio and optical) by damped/red-noise type processes (\citealt{Max2014, Nilsson2018}).

\section{Results and Discussion}\label{res_dis}
In this paper, we analyzed the $\gamma$-ray flaring activity for the blazar PKS 0346-27 during the period 2019 January-2021 December (MJD 58484-59575). Most of the time during this slot this source was in a high state. We also analyzed the follow-up observations in X-ray and UV bands. The archival data of Swift X-ray, UVOT, and Fermi-gamma ray space telescope allowed us to investigate the multi-frequency SED of the source. We study the possible physical mechanism of the source and its jet parameters through theoretical modeling of the multiwavelength SED during four instances when the source was in a high state in $\gamma$-ray energies. A study of this source for an earlier period has been reported in \cite{angioni2019large}. 

We identify five flaring episodes for the $\gamma$-ray light curve. We fit each flare with Sum-of-Exponentials (SoE). The individual flares 1,2 and 4 are again subdivided into two parts (a and b) for a better fit. The corresponding reduced $\chi^2$ values have a range of 0.9-1.8. A total of 39 peaks were modeled and the corresponding rise and decay times were estimated. Our study shows the presence of fast variability during the brightest phase of flaring activity. We found that the fast rise and decays are more frequent which constrains the emission region to be very compact. The statistical distribution of the rise and decay times for the $\gamma$-ray light curve is shown in figure \ref{fig:rise_and_decay_time_distribution}. We found all possible scenarios of symmetric and asymmetric peaks. Considering the fact that the variability in the flare is caused by the interaction of the blob (emission region) with the standing shock (\citealt{1979ApJ...232...34B, 1985ApJ...298..114M}). The symmetric flares are expected when the blob passes through the standing shocks and the particle radiates all its energy within the shock region within the light travel time ($\sim$ R/c). Once the blob is moved out of the shock is moving we expect to see an asymmetric flare. This can also be seen as a scenario where the radiative cooling time is longer than the light travel time and as a result, a long-decay flare is observed. This means the radiative cooling time scale which includes synchrotron and IC cooling time scales is longer than the particle injection which is more or less instantaneous. Symmetric and Asymmetric flares are commonly seen in many blazars (\citealt{2012ApJ...749..191C}).

We have estimated the fast variability time in all the emissions which constrains the size and location of the emission region down the jet. In X-ray, the fastest variability time is found to be 3.24$\pm$0.26 days followed by $\gamma$-ray 1.34$\pm$0.30 and optical-UV with an order of $\sim$0.1 days. Taking the variability of $\gamma$-ray as a reference the size of the emission region is estimated as $\sim 1.05 \times 10^{17}$ cm. From the SED modeling, the best location of the emission region was found to be, R$_H$ = 1.0$\times$10$^{17}$ cm which suggests the emission region is closer to the disk as the BLR starts from $>$10$^{17}$ cm.

We obtained the $\gamma$-ray SEDs for each flare (figure \ref{fig_gamma_sed}). The $\gamma$-ray SEDs were fitted with Power Law, Log-Parabola, Broken Power Law, and Power Law with exponential cutoff models. The fitting parameters obtained from this study are shown in Table \ref{tab:gamma_ray_sed_param}.
In the $\gamma$-ray SED, we noticed a clear cut-off around 15-20 GeV (Table \ref{tab:gamma_ray_sed_param}) and it could be due to $\gamma-\gamma$ absorption by the BLR photons (\citealt{Liu2006}). Combining the information of fast variability and the location of the emission region, it is very much clear that the single-zone emission model would be the best choice to go ahead with. A single-zone leptonic emission model is a well-accepted and widely used scenario that has been used for many FSRQs type blazars (\citealt{angioni2019large, Abdo2010}).

We model the multi-frequency SED of the source from $\gamma$-ray data of Fermi-LAT, X-ray data of Swift-XRT, and UV {and optical} data of Swift-UVOT for flares 1, 2, 3, and 5.  The SED of all flares are well fitted with synchrotron, SSC, and EC from Disk and BLR. The EC-BLR appears to be sub-dominant compared to the EC-disk component. The direct disk thermal emission is also plotted along with the synchrotron and EC components in Figure \ref{fig_broadband_sed_fit}. We have chosen a Power-law with a cutoff as particle distribution and a spherical blob of size, R' which is moving down the jet with Lorentz factor, $\Gamma$. The inner and outer radius of the BLR along with the location of the blob is fixed to a typical value during the modeling. The Doppler factor was also fixed at, $\delta_D$ = 60 taken from \citet{angioni2019large}. The parameters related to particle distribution such as minimum and maximum energy of the particles, cut-off energy, and power-law slope before cut-off, and the parameters related to the blob such as magnetic field and its size are kept free and optimized to the best value through the SED modeling. The BLR optical depth ($\tau_{BLR}$) is another important parameter that is optimized to the best value for all the flares. The $\tau_{BLR}$ is found to be below $30\%$ suggesting weak pair-production or $\gamma-\gamma$ absorption. We fixed the disk luminosity ($L_{Disk}$) and optimized the temperature for all four flares. From the $L_{Disk}$ value, $R_{BLR_{in}}$ is estimated using a scaling relation mentioned in section \ref{sec4}.

A clear thermal signature from the accretion disk was also obtained in \cite{angioni2019large}. The Broken Power Law was used for the electron distribution by \cite{angioni2019large} for the SED modeling while a Power Law with Cutoff is used in this analysis. The low energy spectral slope ($p$) has a range of 2.0-2.3 in 2019 and is similar to the current model where the range is 2.09-2.22. The estimated magnetic field in the 2019 model has a range of 0.82-2.6 in the flaring state compared to the range of 0.56-1.27 in the current model. Both the models have similar order values for $\gamma_{min}$ and $\gamma_{max}$. The power in each component was estimated and compared within the different states. Particles carry more power than magnetic fields suggesting a kinetic-dominated jet. Flare-1 carries the highest power followed by the other flares. The total jet power is estimated to be of the same order of magnitude over all the flares of the current model compared to values in the high state after flare A of \cite{angioni2019large} but are well within the Eddington luminosity of the source. The best fit disk temperature among the different flares is between $10^5 - 10^6$ K which is commonly seen in many FSRQs (\citealt{2018A&A...610A...1M}). The optimized value of the size of the emission region is found to be $\sim 10^{15}$ cm which is consistent with the observational upper limit estimate from the $\gamma$-ray variability time. One thing to note is the \citet{angioni2019large} studied only a small part of the flare between 2017-2019. However, the major flaring happened after that period which is presented in our work. As we know the blazars are highly variable and the flaring epochs of a source do not correlate with each other most of the time due to its stochastic nature therefore, it is less likely that we could derive the exact SED modeling parameters found in \citet{angioni2019large}. In their work, they have fitted the gamma-ray part of the SED with the external Compton from the dusty torus, and sometimes the X-ray is well-fitted with SSC. They also noticed a shift in the synchrotron peak, as well as the EC peak towards higher frequency as the source, goes from a low state to a high flux state. In their case, in the low-state source behave as a low-synchrotron peaked (LSP) blazar and in the high state as an intermediate synchrotron peaked (ISP) blazar. They also suggest it could possibly be a "masquerading" BL Lac in its high state. However, in our study, we only modeled the flaring part so we do not notice any difference or shift in the synchrotron peak.
In all cases, we have found that the synchrotron peak is in the range $10^{14-15}Hz$ as expected for ISPs, consistent with \citet{angioni2019large} at a high state.
In our SED modeling, the X-ray is also well-fitted with SSC but for high-energy peaks, external Compton from the accretion disk was found to be the best scenario.

We have also produced the power spectral density for this source and a power law seems to produce the best fit with a slope of 2.15$\pm$0.87 suggesting variability in this source is dominated by a stochastic process.  Detailed $\gamma$-ray PSD of FSRQ and BL Lacs are studied in \citet{Abdo2010psd} where they have found the power law as the best fit to their PSD without any break. The average PSD slope for a sample of FSRQ was derived as 1.4$\pm$0.1 which is much harder than the PSD slope of individual source PKS 0346-27. The $\gamma$-ray PSD has also been done for many other blazars in past (\citealt{2010ApJ...721.1383A, 2013ApJ...773..177N, 2014ApJ...786..143S, 2015MNRAS.452.1280R, 2017ApJ...849..138K, Goyal_2022}) and most of the time it is well fitted with single power-law. \citet{2013ApJ...773..177N} investigated the 15 AGN and found that 3C 454.3 show a characteristic time scale of 6.8$\times$10$^5$ s and suggested an internal shock model for the variability. Similarly, \citet{2017ApJ...849..138K} studied four Fermi-LAT sources including FRI radio galaxy NGC 1275, BL Lac Mrk 421, FSRQs B2 1520+31, and PKS 1510-089 and found a presence of a break which they argued that their results are broadly consistent with the statistical properties of the magnetic reconnection powered mini jets-in-a-jet model.  

The hard $\gamma$-ray spectrum of PKS 0346-27 could make it a potential target for ground-based imaging atmospheric Cherenkov telescopes. Recently, HESS has reported \footnote{\url{https://www.astronomerstelegram.org/?read=15020/}} to have detected this source with 5$\sigma$ significance. Thus it is the highest redshift blazar surpassing the blazar S3 0218+35 (z = 0.944, \cite{ahnen2016detection}) which is gravitationally lensed as well as surpassing the blazar PKS 1441+25 (z = 0.940, \cite{ahnen2015very}). 
FSRQs are difficult to detect during quiescent states, but because of the large flux variation during flaring states, they can be detected easily. CTA with an order of magnitude better sensitivity and with long exposure will be able to detect even during quiescent state at higher energies. That will provide an opportunity to compare emission mechanisms during the low and high states, as they are expected to be different.
Therefore this source will be one interesting candidate for the southern array of the upcoming Cherenkov telescope array (CTA)\footnote{\url{https://www.cta-observatory.org/}} \citep{bose2022}. Future observations at VHE will be crucial for Extra-galactic Background Light (EBL) studies.

Fermi-LAT has been an important tool for observing high-energy blazars. In the future, it will play a crucial role in order to observe more high redshift $(z \ge 1)$ blazars of the GeV energy range. Continued monitoring of the GeV sky will help to establish the duty cycles of $\gamma$-ray activities in relativistic jets. Moreover, Fermi-LAT is an invaluable tool in order to help ground-based $\gamma$-ray observatories such as Cherenkov telescopes to point their telescopes in the appropriate direction of the source. It will serve the same for the upcoming CTA. Therefore, continued observation of Fermi-LAT in the CTA era would be a great tool to better understand the physics of blazar jets, both in local and high redshift $z \ge 1$ universe.

\section*{Acknowledgements}
We thank the anonymous referee for constructive comments and suggestions which have helped to improve the scientific merit of the work. S. Pramanick acknowledges the support of \href{https://online-inspire.gov.in/}{DST-INSPIRE} Scholarship and Prime Minister's Research Fellowship (\href{https://www.pmrf.in/}{PMRF}). R. Prince is grateful for the support of the Polish Funding Agency National Science Centre, project 2017/26/A/ST9/-00756 (MAESTRO 9) and MNiSW grant DIR/WK/2018/12 and European Research Council (ERC) under the European Union’s Horizon 2020 research and innovation program (grant agreement No. [951549]). D. Bose acknowledges the support of Ramanujan Fellowship-SB/S2/RJN-038/2017. This work made use of Fermi telescope data and the Fermitool package. This work also made use of publicly available packages JetSet and PSRESP. We used a Fermi-user-contributed tool \href{https://fermi.gsfc.nasa.gov/ssc/data/analysis/user/likeSED.py}{likeSED.py} for figure \ref{fig_gamma_sed}.

\section*{Data Availability}
For this work, we have used data from Fermi-LAT, Swift-XRT, and Swift-UVOT telescopes. All the data are available in the public domain. Details are given in section 2. 

\bibliographystyle{mnras}
\bibliography{blazar}


\appendix
\section{Multi-wavelength Variability plots}\label{appendix_sec1}

\begin{figure*}
    \centering
    \includegraphics[width=\textwidth]{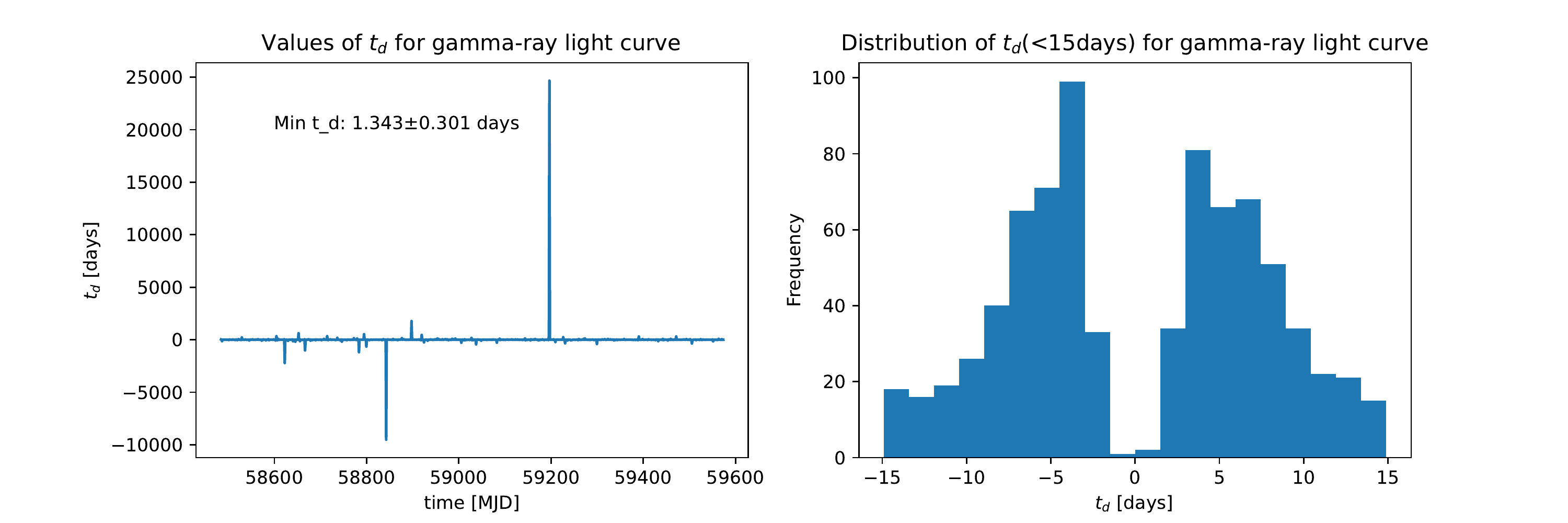}
    \caption{Variablility time values and distribution for $\mathrm{\gamma-ray}$ light curve.}
    \label{fig:gamma_ray_t_d_dist}
\end{figure*}

\begin{figure*}
    \centering
    \includegraphics[width=\textwidth]{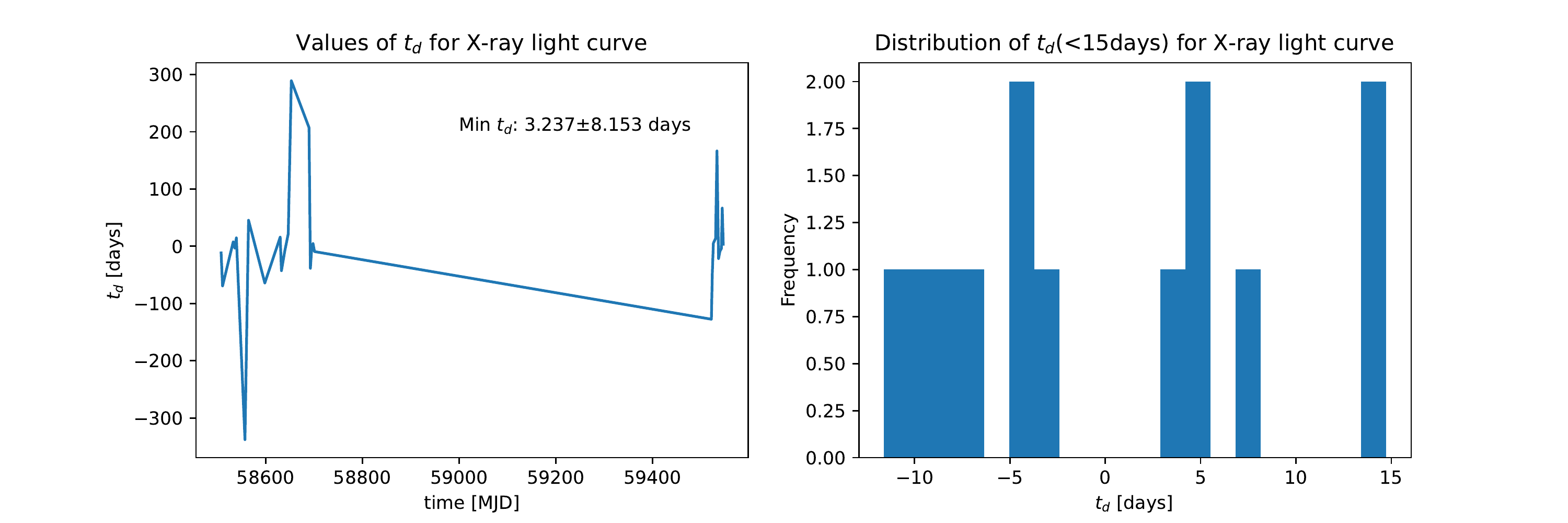}
    \caption{Variablility time values and distribution for X-ray light curve.}
    \label{fig:x_ray_t_d_dist}
\end{figure*}

\begin{figure*}
    \centering
    \includegraphics[width=0.7\textwidth]{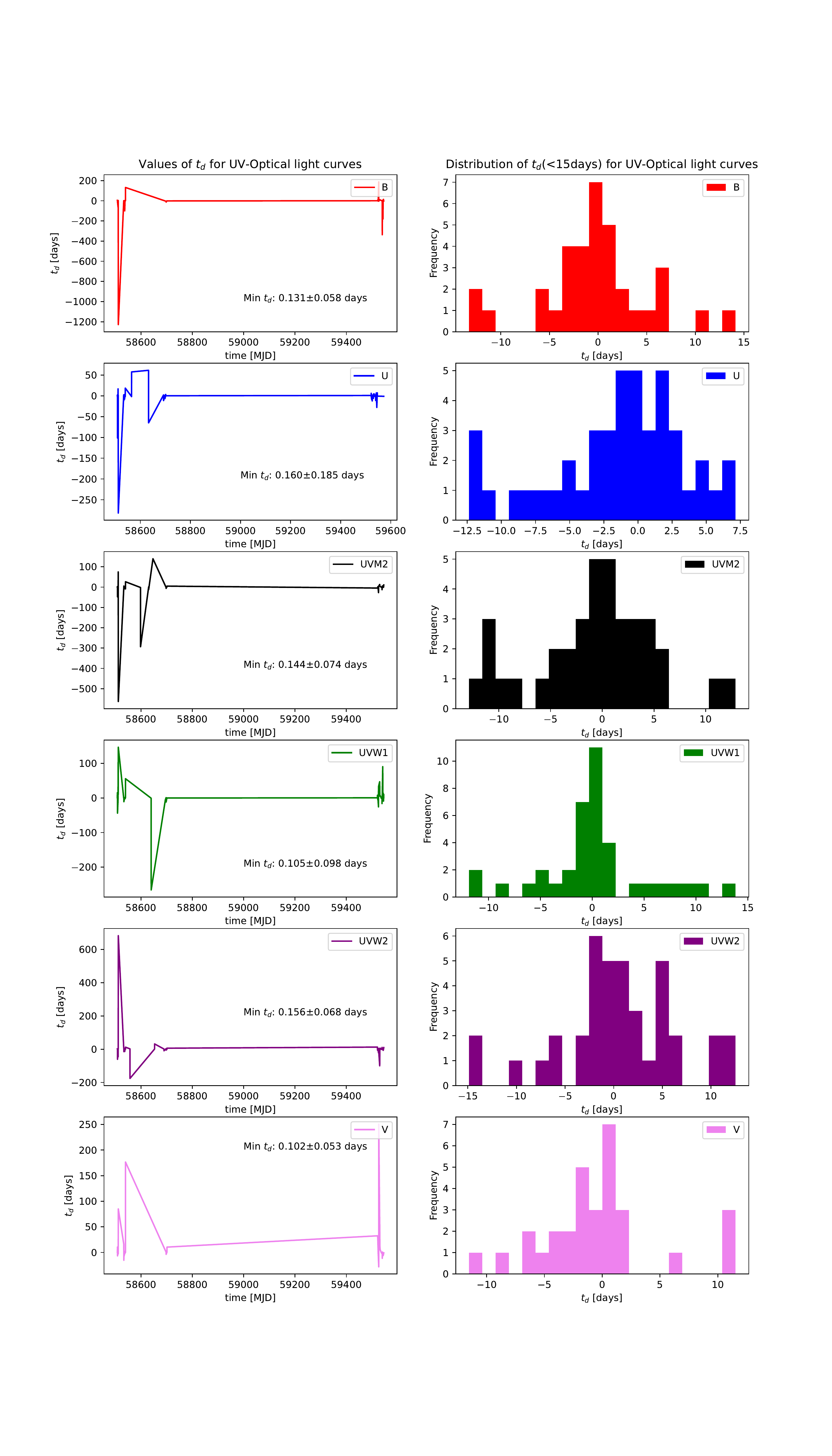}
    \caption{Variablility time values and distribution for UV and Optical light curves.}
    \label{fig:uvot_t_d_dist}
\end{figure*}

\bsp	
\label{lastpage}
\end{document}